\documentclass[10pt,aps,prd,showpacs,nofootinbib,twocolumn]{revtex4-1}
\usepackage{amsfonts,amssymb,amsmath,subfigure,graphicx,epstopdf,bm}
\usepackage[usenames, dvipsnames]{color}
\usepackage[bookmarks]{hyperref}

\newcommand{\Real}{\mathbb{R}}

\date{\today}

\begin{document}

\title{Highly compact neutron stars in scalar-tensor theories of gravity: Spontaneous scalarization versus gravitational collapse}

\author{Raissa F.\ P.\ Mendes}
\email{rmendes@uoguelph.ca}
\affiliation{Department of Physics, University of Guelph, Guelph, Ontario, N1G 2W1, Canada}
\author{N\'{e}stor Ortiz}
\email{nortiz@perimeterinstitute.ca}
\affiliation{Perimeter Institute for Theoretical Physics, 31 Caroline Street, Waterloo, Ontario N2L 2Y5, Canada}

\begin{abstract}
Scalar-tensor theories of gravity are extensions of general relativity (GR) including an extra, nonminimally coupled scalar degree of freedom. A wide class of these theories, albeit indistinguishable from GR in the weak field regime, predicts a radically different phenomenology for neutron stars, due to a nonperturbative, strong-field effect referred to as spontaneous scalarization. This effect is known to occur in theories where the effective linear coupling $\beta_0$ between the scalar and matter fields is sufficiently negative, i.e.~$\beta_0 \lesssim -4.35$, and has been strongly constrained by pulsar timing observations.

In the test-field approximation, spontaneous scalarization manifests itself as a tachyonic-like instability. Recently, it was argued that, in theories where $\beta_0>0$, a similar instability would be triggered by sufficiently compact neutron stars obeying realistic equations of state. In this work we investigate the end state of this instability for some representative coupling functions with $\beta_0>0$. This is done both through an energy balance analysis of the existing equilibrium configurations, and by numerically determining the nonlinear Cauchy development of unstable initial data. We find that, contrary to the $\beta_0<0$ case, the final state of the instability is highly sensitive to the details of the coupling function, varying from gravitational collapse to spontaneous scalarization. In particular,  we show, for the first time, that spontaneous scalarization can happen in theories with $\beta_0>0$, which could give rise to novel astrophysical tests of the theory of gravity. 
\end{abstract}


\maketitle

\section{Introduction}
\label{sec:intro}

Scalar-tensor theories of gravity (STTs) are among the most studied extensions of general relativity (see e.g.~Refs.~\cite{Damour1992,Fujii2003} for reviews). 
These are well-posed theories \cite{Salgado2006} with fruitful applications in cosmology, both in models of cosmic inflation and the dark sector \cite{Faraoni2004,Clifton2012}.
Moreover, a large subset of STTs has the interesting property of eluding tests in the weak field regime, but still predicts radically different strong-field phenomena, particularly for neutron stars. Therefore, STTs provide a test bed to probe deviations from general relativity in its nonperturbative regime.

More concretely, the class of theories that we will consider in this work includes a single scalar field $\phi$ with no self-coupling. Multiscalar generalizations are studied in Refs.~\cite{Damour1992,Horbatsch2015} and the effects of a mass term are discussed in Ref.~\cite{Ramazanoglu2016}.
In local coordinates $\{ x^\mu \}$ defined on a 4-dimensional spacetime $({\cal M},g_{\mu\nu})$, this class of theories is defined by the following action written in geometric units:
\begin{align}
	S[g_{\mu\nu};\phi;\Psi_m ] &= \frac{1}{16\pi} \int{d^4 x \sqrt{-g} \left( R 
	- 2 \nabla_\mu \phi \nabla^\mu \phi \right)} \nonumber \\
	&+ S_m[\Psi_m ; a(\phi)^2 g_{\mu\nu} ],
\label{general_action}
\end{align}
where $g:=\det(g_{\mu\nu})$, $R$ is the Ricci scalar, and the arbitrary function of the scalar field $a(\phi)$ fixes a particular theory.
In this formulation (the so-called {\it Einstein frame}), the scalar field couples minimally to the metric, while matter fields, denoted collectively by $\Psi_m$, couple universally to the conformally rescaled (Jordan) metric $\tilde{g}_{\mu\nu} := a(\phi)^2 g_{\mu\nu}$. Alternatively, action (\ref{general_action}) can be rewritten in terms of the metric $\tilde{g}_{\mu\nu}$, in which case the Einstein-Hilbert term becomes $a(\phi)^{-2} \sqrt{-\tilde{g}} \tilde{R}$, plus additional terms involving derivatives of $\phi$. In this \textit{Jordan frame} formulation, there is a nonminimal coupling of the scalar field to the tensor sector, but there is no explicit coupling to the matter sector.

In STTs, the cosmological evolution would set a constant background value for the scalar field at present time, $\phi_0 := \phi(\tau_0)$, which would be locally modified by the presence of matter inhomogeneities~\cite{Damour1993a,Damour1993b}. Therefore, it is convenient to consider the expansion of the coupling function $a(\phi)$, or, as frequently done, of its logarithmic derivative $\alpha(\phi):=d \ln a(\phi)/d\phi$, around the background scalar field:
\begin{equation}\label{eq:a_taylor}
	\alpha(\phi) = \alpha_0 + \beta_0 (\phi - \phi_0) + O[(\phi-\phi_0)^2].
\end{equation}
When only the first term in this expansion is present, the theory reduces to the Jordan-Brans-Dicke proposal \cite{Jordan1959,Brans1961}.
This class of theories is, however, considerably constrained by solar system experiments, which enforce $|\alpha_0|$ to be very small, namely $|\alpha_0| \lesssim 3.4 \times 10^{-3}$ \cite{Bertotti2003}. On the other hand, solar system experiments are not as severe in constraining the coefficients of the higher order terms in Eq.~(\ref{eq:a_taylor}). Indeed, in the particular case of $\alpha_0=0$, these experiments cannot distinguish a STT from GR at all, since both agree to all post-Newtonian orders independently of $\{\beta_0,...\}$ \cite{Damour1996a}.

The second term in Eq.~(\ref{eq:a_taylor}) is responsible for the most striking effects in the strong-field regime of STTs. If $\beta_0$ is sufficiently negative, i.e. $\beta_0 \lesssim -4.35$, compact stars are predicted to undergo a transition from a solution close to GR (or even identical, if $\alpha_0=0$) to a configuration with a nontrivial scalar field profile and non-negligible scalar charge. This effect, known as {\it spontaneous scalarization} \cite{Damour1993,Damour1996,Salgado1998,Harada1998}, 
opened a new avenue to test these theories, e.g., based on observations of neutron stars in binary systems, which would lose energy faster due to gravitational-wave emission in the extra scalar channel.
The lack of such an effect in pulsar timing data has been used to exclude the values $\beta_0\lesssim -4.5$ \cite{Freire2012}, thereby considerably restricting the parameter space where spontaneous scalarization was predicted. 
With the recent detection of gravitational waves by Advanced LIGO \cite{Abbott2016a} and the beginning of the era of gravitational-wave astronomy, the presence of this effect in dynamical settings \cite{Barausse2013,Shibata2014,Palenzuela2014,Sampson2014} can also potentially be tested.

Due to the existence of the scalarization effect in theories with $\beta_0 < 0$, works in the past three decades have focused almost exclusively in this region of the parameter space (see \cite{Berti2015a} for a broad review). 
However, it was recently realized \cite{Mendes2015} that many realistic equations of state for nuclear matter can support stars which are compact enough to exhibit a similar effect in theories with $\beta_0>0$.
In order to demonstrate this, Ref.~\cite{Mendes2015} mainly explored the test-field approximation, in which the scalarization effect manifests itself as a tachyonic-like instability. It was left as an open question whether the end state of this instability would be a stable, scalarized solution (as in the $\beta_0<0$ case), or would lead to a different outcome, such as an explosion or gravitational collapse. The first numerical simulations addressing this issue were reported in Ref.~\cite{Palenzuela2016}. The authors restricted their attention to the model obtained by truncating the series expansion of Eq.~(\ref{eq:a_taylor}) up to linear order---which has been widely adopted in the literature since the seminal work by Damour and Esposito-Far\`{e}se \cite{Damour1996}---, and concluded that the end state of the instability would be, in general, collapse to a black hole.

The aim of this work is to perform a more detailed analysis of the end state of the instability of highly compact neutron stars in STTs with $\beta_0>0$ and investigate its astrophysical implications. We consider two different models with representative coupling functions, which coincide up to the linear term in Eq.~(\ref{eq:a_taylor}). One is again the Damour-Esposito-Far\`ese (DEF) model obtained by neglecting quadratic and higher-order terms in Eq.~(\ref{eq:a_taylor}); the other is an analytical approximation to the physically interesting case of a scalar field that couples nonminimally to gravity, by means of a $\xi \tilde{R} \phi^2$ interaction term in the Jordan frame action, where $\xi \in \Real$. 
Our analysis follows two main routes. First, we construct static equilibrium solutions in theories with $\beta_0>0$, and determine whether scalarized solutions can be energetically favored over the unstable GR-like configurations, and thus be a plausible end state of their evolution. Second, we numerically solve the initial value problem for the scalar-tensor-fluid evolution equations in order to determine the stability of the various equilibrium stellar configurations, and dynamically investigate the final state of the unstable ones. The conclusions we draw from both approaches are complementary and in complete agreement with each other.

The most intriguing finding of our work is the manifest model dependence of the results. In fact, within the DEF model we find that the existing scalarized solutions are all unstable and energetically disfavored with respect to the GR-like solution and that, when the latter is unstable, it undergoes gravitational collapse---in agreement with Ref.~\cite{Palenzuela2016}. On the other hand, within the model that mimics nonminimally coupled fields, we find the existence of stable, energetically favored scalarized solutions, which are numerically seen to be the end state of the instability in many cases. To the best of our knowledge, this is the first dynamical demonstration that spontaneous scalarization can take place in STTs with $\beta_0>0$.

This model dependence when $\beta_0>0$ contrasts sharply with the $\beta_0<0$ case, in which higher order terms in Eq.~(\ref{eq:a_taylor}) are known to influence only quantitatively the properties of scalarized solutions.
This feature of the $\beta_0>0$ case is, on the one hand, less ``convenient'', in the sense that eventual observational constraints cannot be expressed as generic bounds on $\beta_0$, but must be attributed to particular classes of models. On the other hand, it also means that observations could potentially be used to probe deeper into the structure of the coupling function $\alpha(\phi)$, due to the richer phenomenology present in this case.

The paper is organized as follows. In Sec.~\ref{sec:field_equations} we present the equations governing the dynamics of STTs in spherical symmetry and write them in a flux-conservative form suitable to our numerical techniques. The static limit of these equations is also discussed, as well as the numerical algorithm to construct equilibrium solutions. In Sec.~\ref{sec:physical_setup} we describe our chosen equation of state, and define the two forms for the coupling function that are used in what follows. In order to compare our results with previous analyses in a relatively self-contained manner, in Sec.~\ref{sec:previous} we describe (i) the test-field approximation and the appearance of unstable scalar modes in GR-like configurations, and (ii) properties of scalarized solutions when $\beta_0<0$, as well as illustrative results from our numerical simulations. Section \ref{sec:betaplus} contains our main results. We conclude in Sec.~\ref{sec:discussion} with further discussions. Details on numerical methods and self-convergence tests are deferred to the Appendix. We use units such that $c = 1 = G$ throughout the text.

\section{Field equations}
\label{sec:field_equations}

The field equations obtained from varying the action in  Eq.~(\ref{general_action}) with respect to the metric $g_{\mu\nu}$ and the scalar field $\phi$ in local coordinates $\{ x^\mu\}$ read
\begin{eqnarray}
	G_{\mu\nu} - 2 \nabla_\mu \phi \nabla_\nu \phi + 
	g_{\mu\nu} \nabla_\rho \phi 
	\nabla^\rho \phi  &=& 8\pi a^2 \tilde{T}_{\mu\nu} ,
	\label{eq:G_eq} \\
	\nabla^\mu \nabla_
	\mu \phi &=& -4 \pi a^4 \alpha \tilde{T},
	\label{eq:phi_eq}
\end{eqnarray} 
where
\begin{equation} \label{eq:alpha_def}
\alpha(\phi) := \frac{d\ln a(\phi)}{d\phi},
\end{equation}
$\tilde{T} := \tilde{g}_{\mu \nu }\tilde{T}^{\mu \nu}$, and $\tilde{T}^{\mu\nu} := 2 (-\tilde{g})^{-1/2} \delta S_m [\Psi_m ; \tilde{g}_{\rho\sigma} ]/\delta \tilde{g}_{\mu\nu}$ is the stress-energy-momentum tensor of the matter fields, which is covariantly conserved in the sense that
\begin{equation} \label{eq:eom}
\tilde{\nabla}_\nu \tilde{T}^{\mu\nu} = 0,
\end{equation}
where $\tilde{\nabla}$ is the covariant derivative compatible with the Jordan metric $\tilde{g}_{\mu\nu}= a(\phi)^2 g_{\mu\nu}$. Note that quantities with a tilde are associated to the Jordan metric, which also lowers and raises their indices. In this work, we choose to evolve the Einstein frame metric due to the simplicity of the resulting field equations, but we describe the fluid in the Jordan frame, in which the equations of motion~(\ref{eq:eom}) have a more natural interpretation. For a discussion on the equivalence of Einstein and Jordan frames, see e.g.~Refs.~\cite{Flanagan2004,Sotiriou2008}.
In this work, we model neutron stars by spherically symmetric perfect fluids with stress-energy-momentum tensor given by
\begin{equation} \label{eq:Tmunu}
\tilde{T}^{\mu\nu} = \tilde{\epsilon} \tilde{u}^\mu \tilde{u}^\nu + \tilde{p} (\tilde{g}^{\mu\nu} + \tilde{u}^\mu \tilde{u}^\nu),
\end{equation}
where $\tilde{\epsilon}$ and $\tilde{p}$ are the fluid's total energy density and pressure, respectively, as measured by observers comoving with the fluid elements, whose 4-velocity $\tilde{\bf u}$ is normalized according to $\tilde{g}_{\mu \nu} \tilde{u}^\mu \tilde{u}^\nu = -1$.
The description of the star is completed by specifying a cold equation of state, $\tilde{p}=\tilde{p}(\tilde{\rho})$, where $\tilde{\rho}$ is the fluid's baryon mass density as measured by observers comoving with the fluid, and defines the baryon mass current $\tilde{\bf J} := \tilde{\rho} \tilde{\bf u}$, which is locally conserved according to
\begin{equation} \label{eq:MB_conserv}
\tilde{\nabla}_\mu \tilde{J}^\mu = 0.
\end{equation}

Our particular choice of equation of state will be described and justified in Sec.~\ref{sec:eos}.

\subsection{Evolution equations in spherical symmetry}

We are interested in finding solutions of the initial value problem for the relativistic hydrodynamic system consisting of the scalar-tensor field equations~(\ref{eq:G_eq})~and~(\ref{eq:phi_eq}) coupled to the Euler equations~(\ref{eq:eom})~and~(\ref{eq:MB_conserv}).
To this purpose, we follow the standard $3+1$ formalism~\cite{Alcubierre-Book}, which splits the 4-dimensional spacetime into 3-dimensional Cauchy hypersurfaces labeled by a {\it coordinate time} $t$. We assume spherical symmetry and thus foliate the spatial hypersurfaces in 2-spheres described in the usual spherical coordinates $\{ r, \vartheta, \varphi \}$, so the spacetime coordinate basis is $\{ \partial_t, \partial_r, \partial_\vartheta, \partial_\varphi \}$. Let $\tilde{\bf n}$ be a future directed timelike unit vector field orthogonal to the $t = \textrm{cnt}$ hypersurfaces, meaning $\tilde{n}^\mu \tilde{n}_\mu = -1$, and $\tilde{n}^\mu \tilde{e}_{(j) \mu} = 0$ for $\tilde{\bf e}_{(j)} := \partial_j/a$, $j \in \{ r,\vartheta,\varphi \}$.
Demanding zero spatial shift between the Cauchy hypersurfaces, which is known as {\it polar slicing condition}, the relation $\partial_t = aN(t,r) \tilde{\bf n}$ defines the {\it lapse function} $N$. Furthermore, we impose the {\it radial gauge}, which consists in the coincidence of the coordinate $r$ with the areal radius of the 2-spheres. The polar slicing condition, together with the radial gauge, imply the following form of the spacetime line element in the Einstein frame:
\begin{equation}
ds^2 = - N(t,r)^2 dt^2 + A(t,r)^2 dr^2 + r^2 
(d\vartheta^2 + \sin^2\vartheta d\varphi^2).
\end{equation}
In analogy with the static, vacuum case, the radial metric function $A$ is written in terms of an auxiliary {\it mass aspect function} $m$ through $A(t,r) := [1-2 m(t,r)/r ]^{-1/2}$.

Particularly relevant is the family of observers moving along the integral curves of the vector field $\tilde{\bf n}$, known as {\it Eulerian observers}. Their relation to the ones comoving with the fluid is given by the Lorentz factor $\Gamma := -\tilde{n}_\mu \tilde{u}^\mu = a N \tilde{u}^t$, which by virtue of the normalization $\tilde{u}_\mu\tilde{u}^\mu = -1$, can be written as
\begin{equation}
\Gamma = \left(1 - A^2 v^2 \right)^{-1/2},
\end{equation}
where $A v := (A/N)(dr/dt)$ is the fluid's radial velocity as measured by an Eulerian observer.

At this point, we could proceed to project the fluid equations~(\ref{eq:eom})~and~(\ref{eq:MB_conserv}) along the basis $\{ \tilde{\bf n}, \tilde{\bf e}_{(j)} \}$ adapted to Eulerian observers, write down evolution equations for the set of {\it primitive variables} $\{ \tilde{\epsilon}, v, \tilde{p} \}$, and attempt to solve them by a standard finite differences numerical scheme. However, the hydrodynamic equations are known to generically develop shocks and rarefaction waves characterized by unbounded gradients in the fluid quantities, which standard finite differences methods are unable to handle~\cite{LeVeque-Book_FD}. Instead, we implement a finite volume numerical scheme together with a high resolution shock capturing (HRSC) method designed to consistently treat rarefaction and shock propagation (see the Appendix for details). In particular, finite volume methods require the evolution equations to be written as a hyperbolic system of conservation laws~\cite{LeVeque-Book_FV}, which in a spherically symmetric spacetime takes the form
\begin{equation}\label{eq:flux-conservative-form}
\frac{\partial}{\partial t}(A {\bf q}) + \frac{1}{r^2} \frac{\partial}{\partial r}\left(NA r^2 {\bf F}({\bf q})\right) = {\bf S}({\bf q}),
\end{equation}
where ${\bf q}$ is a state vector of conserved quantities with associated flux and source vectors ${\bf F}({\bf q})$ and {\bf S}({\bf q}), respectively. To that end, let us construct alternative fluid variables all measured by Eulerian observers, namely the total energy density $\tilde{E}$, the baryon mass density $\tilde{D}$, the radial momentum density $\tilde{S}$, and the internal energy density $\tilde{\tau}$, defined by
\begin{eqnarray}
\tilde{E} &:=& \tilde{T}^{\mu \nu} \tilde{n}_\mu \tilde{n}_\nu = \Gamma^2 (\tilde{\epsilon} + \tilde{p}) - \tilde{p}, \label{eq:E} \\
\tilde{D} &:=& - \tilde{J}^\mu \tilde{n}_\mu = \tilde{\rho} \Gamma, \label{eq:D} \\
\tilde{S} &:=& -\tilde{T}^{\mu \nu} \tilde{n}_\mu \tilde{e}_{(r) \nu} = (\tilde{E} + \tilde{p}) A^2 v , \label{eq:S}\\
\tilde{\tau} &:=& \tilde{E} - \tilde{D}. \label{eq:tau}
\end{eqnarray}
Also, the wave equation~(\ref{eq:phi_eq}) governing the scalar field dynamics can be split into a system of first-order, inhomogeneous advection equations, which can be put into the flux-conservative form~(\ref{eq:flux-conservative-form}), and thus be solved using the same finite volume scheme as for the fluid equations. To this purpose, we introduce the scalar field variables
\begin{equation}\label{eq:eta_and_psi}
\eta := \frac{1}{A} \frac{\partial \phi}{\partial r}, \qquad
\psi := \frac{1}{N} \frac{\partial \phi}{\partial t}.
\end{equation}
In terms of the conserved quantities~(\ref{eq:D})-(\ref{eq:eta_and_psi}), the scalar field evolution equation~(\ref{eq:phi_eq}) together with the Euler equations~(\ref{eq:eom})~and~(\ref{eq:MB_conserv}), can be collectively written as a flux-conservative system of the form~(\ref{eq:flux-conservative-form}) with ${\bf q} = ( \tilde{D}, \tilde{S}, \tilde{\tau} , \eta, \psi )^T$,
${\bf F} = (F_{\tilde{D}}, F_{\tilde{S}}, F_{\tilde{\tau}}, F_\eta, F_\psi)^T$, and ${\bf S} = (S_{\tilde{D}}, S_{\tilde{S}}, S_{\tilde{\tau}}, S_\eta, S_\psi)^T$, where\footnote{A flux-conservative formulation of the fluid equations in STTs can also be found in Refs.~\cite{Novak2000,Gerosa2016}. Note that their definitions of conserved variables slightly differs form ours.}
\begin{align}
F_{\tilde{D}} & = \tilde{D} v, \label{eq:F_D}\\
F_{\tilde{S}} & =  \tilde{S} v + \tilde{p}, \\
F_{\tilde{\tau}} & = (\tilde{\tau} + \tilde{p}) v, \\ 
F_\eta & = -\psi/A, \\
F_\psi & = -\eta/A,
\end{align}
and
\begin{align}
S_{\tilde{D}} =& -3 \alpha N A \tilde{D} \left( \psi + Av \eta \right), \\
S_{\tilde{S}} =& 2 N A \frac{\tilde{p}}{r} 
	- NA^3 \frac{m}{r^2} \left( \tilde{D} + \tilde{\tau} + \tilde{S}v + \tilde{p} \right) \nonumber \\
	& - 4 \alpha NA \psi \tilde{S} -\alpha NA^2 \eta \left( \tilde{D} + \tilde{\tau} + 3 \tilde{S} v + \tilde{p} \right) \nonumber \\
	& - \frac{1}{2} NA^3 r \left( \eta^2+\psi^2 \right) 
	\left( \tilde{D} + \tilde{\tau} - \tilde{S} v - \tilde{p} \right), \\
S_{\tilde{\tau}} =& -NA \frac{m}{r^2} \tilde{S} - \alpha NA \psi 
	\left( 3 \tilde{\tau} + \tilde{S} v + 3\tilde{p} \right) \nonumber \\
	& - \alpha NA^2 \eta v \left( \tilde{D} + 4 \tilde{\tau} + 4 \tilde{p} \right) -NA^2 r \psi \eta \left( \tilde{S} v + \tilde{p}\right) \nonumber \\
	&- \frac{1}{2} NA r \left(\eta^2+\psi^2 \right) \tilde{S}, \\ 
S_\eta =& -2N \frac{\psi}{r}, \\
S_\psi =& - 4\pi \alpha a^4 N A \left( \tilde{D} + \tilde{\tau} - \tilde{S} v - 3 \tilde{p} \right).\label{eq:S_psi}
\end{align}

Regarding the spherically symmetric spacetime evolution within the $3+1$ formalism, it can be fully determined from the fluid and scalar field data at each time step by integrating the {\it Hamiltonian constraint} and the {\it lapse condition}
\begin{align}
\frac{\partial m}{\partial r} &= \frac{r^2}{2} \left[ \eta^2 + \psi^2 + 8\pi a^4 \left( \tilde{\tau} + \tilde{D} \right) \right], \label{eq:Hamiltonian}
\\
\frac{\partial N}{\partial r} &= A^2 N \left[ \frac{m}{r^2} + 4\pi r a^4 \left( \tilde{p} + \tilde{S} v \right) + \frac{r}{2} \left( \eta^2+\psi^2 \right) \right],\label{eq:lapse_condition}
\end{align}
respectively. The equation resulting from the {\it momentum constraint},
\begin{equation}\label{eq:momentum_constraint}
\frac{\partial m}{\partial t} = r^2 \frac{N}{A^2} \left( A \eta\psi - 4\pi a^4 \tilde{S} \right),
\end{equation}
which overdetermines the system, is usually discarded. However, we find it convenient to evolve the metric function $m(t,r)$ through Eq.~(\ref{eq:momentum_constraint}), and employ the Hamiltonian constraint~(\ref{eq:Hamiltonian}) as a natural monitor of the accuracy and convergence of the numerical solutions. 
An additional test of the numerical code can be performed by monitoring the evolution of the total baryonic mass, defined as
\begin{equation} \label{eq:Total_BM}
M_b = \int_0^{R_s} 4 \pi r^2 \tilde{D}~a(\phi)^3( 1- 2m/r)^{-1/2} dr,
\end{equation}
which is conserved as a consequence of Eq.~(\ref{eq:MB_conserv}). We refer to the Appendix for details on convergence tests.

\subsection{Static limit}

The static limit of the field equations is of particular relevance not only because the initial data for our numerical simulations will consist of static equilibrium solutions, but also because by studying properties of these equilibrium solutions, we can gain much insight into the outcome of the numerical experiments. Therefore, here we briefly describe the static limit of the field equations, the appropriate boundary conditions, and the method employed to solve them.

In the static limit, in terms of the primitive variables, the equations for the metric functions, scalar field, and fluid pressure reduce to 
\begin{align}
 &\frac{d m}{dr} = 4\pi r^2 a^4 \tilde{\epsilon} + \frac{r}{2} (r-2m) \Big(\frac{d\phi}{dr}\Big)^2 \label{eq:dm}\\
 &\frac{d \ln N}{dr} = \frac{4\pi r^2 a^4 \tilde{p}}{r - 2m} +\frac{r}{2} \Big(\frac{d\phi}{dr}\Big)^2 + \frac{m}{r(r-2m)} \label{eq:dn} \\
 &\frac{d^2\phi}{dr^2} = \frac{4\pi r a^4}{r-2m} \! \left[ \alpha (\tilde{\epsilon} - 3\tilde{p}) + r (\tilde{\epsilon} - \tilde{p}) \frac{d\phi}{dr} \right ]\! -\frac{2(r-m)}{r(r-2m)} \frac{d\phi}{dr} \label{eq:dphi} \\
 &\frac{d\tilde{p}}{dr} = -(\tilde{\epsilon} + \tilde{p}) \left[  \frac{4\pi r^2 a^4 \tilde{p}}{r-2m} \! + \! \frac{r}{2} \Big(\frac{d\phi}{dr}\Big)^2 \!\! + \! \frac{m}{r(r-2m)} \! + \! \alpha \frac{d\phi}{dr} \right], \label{eq:dp}
\end{align}
which generalize the Tolman-Oppenheimer-Volkoff equations of hydrostatic equilibrium.
To close the system, an equation of state (EoS) for the fluid must be specified, and our choice is described below in Sec.~\ref{sec:eos}. 

Given values for the asymptotic amplitude $\phi_0$ of the scalar field and the central pressure $\tilde{p}_c$ of the star, Eqs.~(\ref{eq:dm})-(\ref{eq:dp}) can be solved subject to the boundary conditions 
\begin{align}
&m(0) = 0, \quad \lim_{r\to\infty}N(r) = 1, \quad \lim_{r\to\infty}\phi(r) = \phi_0, \nonumber \\
&\frac{d\phi}{dr}(0) = 0, \qquad \tilde{p}(0) = p_c, \qquad \tilde{p}(R_s) = 0, \label{eq:bc}
\end{align}
where the last equation defines the stellar radius $R_s$. This task is simplified by the existence of an analytical solution of Eqs.~(\ref{eq:dm})-(\ref{eq:dphi}) in vacuum \cite{Coquereaux}. Therefore, it suffices to solve the system of equations in the stellar interior and perform a matching to the exterior analytical solution at the stellar radius. In practice, we integrate Eqs.~(\ref{eq:dm})-(\ref{eq:dp}) with a fourth-order Runge-Kutta algorithm starting with the appropriate boundary conditions at $r=0$ [cf.~Eq.~(\ref{eq:bc})], supplemented with a guess $\phi(0) = \phi_c$, and then iterate on $\phi_c$ until the condition \cite{Damour1993}
\begin{equation} \label{match}
\phi_s -\phi_0 + \frac{2 \psi_s}{\sqrt{\dot{\nu}_s^2+4\psi_s^2}} \textrm{arctanh} \left[ \frac{\sqrt{\dot{\nu}_s^2 +4\psi_s^2}}{\dot{\nu}_s +2/R_s} \right] = 0
\end{equation}
is satisfied up to a given numerical accuracy (see the Appendix for details). Here, the subscript $s$ indicates quantities evaluated at $R_s$; also, $\psi_s := (d\phi/dr)_s$ and $\dot{\nu}_s := 2(d\ln N/dr)|_s = R_s \psi_s^2 + 2 m_s/[R_s(R_s-2m_s)]$. Equation (\ref{match}) follows directly from algebraic manipulation of the matching conditions to the exterior solution. It also follows that the ADM mass and the scalar charge of the solution are given by
\begin{align}
M &= \frac{R_s^2 \dot{\nu}_s}{2} \left( 1-\frac{2m_s}{R_s} \right)^\frac{1}{2} \times \nonumber \\
& \exp \left[ \frac{-\dot{\nu}_s}{\sqrt{\dot{\nu}_s^2+4\psi_s^2}} \textrm{arctanh} \left( \frac{\sqrt{\dot{\nu}_s^2+4\psi_s^2}}{\dot{\nu}_s +2/R_s} \right) \right], \\
\omega & = - 2 M \psi_s/\dot{\nu}_s,
\end{align}
respectively, where $\omega$ is defined from the asymptotic behavior of the field at spatial infinity, through $\phi = \phi_0 + \omega/r + \mathcal{O}(1/r^2)$ \cite{Damour1992}.

\section{Physical setup}
\label{sec:physical_setup}
\subsection{Equation of state}
\label{sec:eos}

Isolated neutron stars are characterized to a good approximation by a cold equation of state \cite{Haensel2007}, which encodes relevant information about the nuclear microphysics. Here we will assume a polytropic EoS of the form
\begin{equation}
	\tilde{p}(\tilde{\rho}) = K \rho_0 (\tilde{\rho}/\rho_0)^\gamma,
\end{equation}
where $\rho_0$ is some reference value for the rest-mass density and $K$ and $\gamma$ are dimensionless constants. The energy density is then determined by the first law of thermodynamics, with the result
\begin{equation}
\tilde{\epsilon}(\tilde{\rho}) = \tilde{\rho} + (\gamma-1)^{-1} \tilde{p}.
\end{equation}
Our choice of polytropic parameters is $\gamma = 3$ and $K = 0.005$, and we take $\rho_0 = 1.66 \times 10^{14}$g/cm${}^3$, which is of the order of the nuclear saturation density and is a common choice in the literature. The exponent $\gamma=3$ is a typical effective polytropic parameter for the core of neutron stars, according to realistic equations of state (see e.g.~values for $\Gamma_1$ in Table III of Ref.~\cite{Read2009}). The value of $K$ is chosen in order to guarantee that the model predicts a maximum neutron star mass consistent with observations \cite{Antoniadis2013}: within GR, this is computed to be $2.03 M_\odot$, for a star with central density  $\tilde{\rho}_{c} \simeq 12.9 \rho_0$ and compactness $M/R_s \simeq 0.316$. Also, the mass-radius relation predicted by this EoS is in agreement with astrophysical and experimental constraints (see e.g.~Fig.~10 of Ref.~\cite{Ozel2016}).
Finally, according to this EoS, stars with a compactness larger than $M/R_s \simeq 0.27$ (or, alternatively, $\tilde{\rho}_c \gtrsim 8.94 \rho_0$) have the property that the trace of the energy-momentum tensor of the fluid, $\tilde{T} = 3\tilde{p} - \tilde{\epsilon}$, is positive in a region around the stellar center. This is a crucial property for the effects we analyze in this work, as discussed in Sec.~\ref{sec:stability_GR}.

Of course, this simple EoS cannot be said to be entirely realistic. In particular, it does not match models of the relatively well-understood microphysics below nuclear density \cite{Douchin2001} and allows for superluminal propagation of sound, $v_s = \sqrt{d\tilde{p}/d\tilde{\epsilon}} > 1$ when $\tilde{\rho} \gtrsim 11.55 \rho_0$, i.e., inside stars with $M/R_s \gtrsim 0.305$. However, by suitably modifying the EoS at low densities, these drawbacks could be avoided, while leaving essentially unaltered our main results, which depend strongly only on the core part of the equation of state. This simple choice will therefore be enough for the purposes of this work. See Ref.~\cite{Mendes2015} for an analysis of the EoS dependence of the instability discussed in Sec.~\ref{sec:stability_GR}.

Of course, this simple EoS cannot be said to be entirely realistic. In particular, (i) it does not match models of the relatively well-understood microphysics below nuclear density \cite{Douchin2001} and (ii) allows for superluminal propagation of sound, $v_s = \sqrt{d\tilde{p}/d\tilde{\epsilon}} > 1$ when $\tilde{\rho} \gtrsim 11.55 \rho_0$, i.e., inside stars with $M/R_s \gtrsim 0.305$. However, the drawbacks (i) and (ii) could be avoided by slightly modifying the EoS at low and high densities, respectively. We have verified that our main results remain qualitatively unaltered when a more realistic EoS is adopted. Still, the choice of a simple $\gamma = 3$ polytrope is enough for the purposes of this work. See Ref.~\cite{Mendes2015} for an analysis of the EoS dependence of the instability discussed in Sec.~\ref{sec:stability_GR}.

\subsection{Coupling function}

In the following, we will consider two main representative forms for the function $a(\phi)$ [or, alternatively, its logarithmic derivative $\alpha(\phi)$ defined in Eq.~(\ref{eq:alpha_def})]. They are motivated by the facts that $\alpha_0 := \alpha(\phi_0)$ is constrained by solar system experiments to be very close to zero, $|\alpha_0| \lesssim 3.4 \times 10^{-3}$ \cite{Bertotti2003}, and that the phenomenon of spontaneous scalarization depends crucially on the value of $\beta_0 := \alpha'(\phi_0)$. Specifically, we define:
\begin{align}
\textrm{{\bf Model 1} (M1)}:\quad &a(\phi) = \left[ \cosh \left(\sqrt{3} \beta (\phi-\phi_0)\right) \right]^\frac{1}{3\beta}, \nonumber \\
& \alpha(\phi) = \frac{1}{\sqrt{3}} \tanh [\sqrt{3} \beta (\phi-\phi_0)]; \label{eq:alpha1} \\
\textrm{{\bf Model 2} (M2):}\quad &a(\phi) = e^{\frac{1}{2}\beta (\phi - \phi_0)^2}, \nonumber \\
& \alpha(\phi) = \beta(\phi - \phi_0), \label{eq:alpha2} 
\end{align}
where $\beta \in \Real$.
In both models, we have $\alpha_0=0$ \footnote{We have verified that the introduction of a nonzero value for $\alpha_0$, but still compatible with current constraints, would not affect the conclusions of our work.} and $\beta_0 = \beta$, so that they only differ by higher-order terms in the expansion (\ref{eq:a_taylor}): $\alpha_\textrm{M1}(\phi) - \alpha_\textrm{M2}(\phi) = - \beta^3 (\phi-\phi_0)^3 + O[(\phi-\phi_0)^5]$. 
Note that, in practical computations within these models, the actual value of $\phi_0$ is irrelevant, since it can be absorbed in the definition of the scalar field. 

Model 2 is the most common in the literature of STTs since the works by Damour and Esposito-Far\`ese, providing the simplest coupling function exhibiting the spontaneous scalarization effect they unveiled \cite{Damour1993}. 

The motivation behind Model 1 is to provide an analytical approximation to the coupling function that arises from a more fundamental\footnote{At the classical level it is natural to include the coupling term $\xi \tilde{R} \Phi^2$ in the action when generalizing the flat-space theory to a curved spacetime. From a quantum-field-theory perspective, the inclusion of such a term can be required by the renormalization of the scalar field in a curved background \cite{Birrell1982}.} theory containing a massless scalar field $\Phi$ nonminimally coupled to gravity. The action describing this theory is usually formulated in the Jordan frame as
\begin{align}
	S[\tilde{g}_{\mu\nu}; \Phi; \Psi_m] &= \frac{1}{2} \! \int{d^4 x \sqrt{-\tilde{g}} \left( \frac{\tilde{R}}{8\pi}  - \tilde{\nabla}_\mu \Phi \tilde{\nabla}^\mu \Phi - \xi \tilde{R} \Phi^2 \!\right)}	
	\nonumber \\
	&+ S_m[\Psi_m ; \tilde{g}_{\mu\nu} ],
\label{eq:S_NMC}
\end{align}
with $\xi \in \mathbb{R}$. 
It can be rewritten in the Einstein frame, in the form of Eq.~(\ref{general_action}), by means of a field redefinition $\Phi = \Phi(\phi)$ and a conformal transformation of the metric $\tilde{g}_{\mu\nu} = a^2(\phi)g_{\mu\nu}$, such that
\begin{align}
&\frac{d\phi}{d\Phi} = 2\sqrt{\pi} \frac{\sqrt{1-8\pi \xi(1-6\xi) \Phi^2}}{1-8\pi \xi \Phi^2}, \label{eq:dPhi}\\
&a(\phi) = (1-8\pi\xi\Phi(\phi)^2)^{-1/2}. \label{eq:a_NMC}
\end{align}

From the action (\ref{eq:S_NMC}), an interpretation that loosely follows is that Newton's constant ($G = 1$ in
our convention) gets replaced in this theory by the effective gravitational coupling $G_\textrm{eff} = G a(\phi)^2$, which is enhanced or diminished from its GR value if $\xi>0$ or $\xi<0$, respectively. 
Note also that, for the transformations (\ref{eq:dPhi}) and (\ref{eq:a_NMC}) to be well-defined, we must have $8\pi \xi \Phi^2<1$. For $\xi< 0$, this condition is trivially satisfied. For $\xi > 0$, it implies the existence of a critical value $\Phi_\textrm{cr} = 1/\sqrt{8\pi \xi}$, for which gravity would become ``infinitely attractive'' in the sense that $G_\textrm{eff} = G a(\phi)^2$ diverges as $\Phi \to \Phi_\textrm{cr}$. It is interesting to notice that any legitimate $\Phi \in (-\Phi_\textrm{cr}, \Phi_\textrm{cr})$ is mapped to some $\phi \in (-\infty,\infty)$ by Eq.~(\ref{eq:dPhi}), so that no restriction to the scalar field values exists in the Einstein frame description.

\begin{figure}[t]
\includegraphics[width=8.5cm]{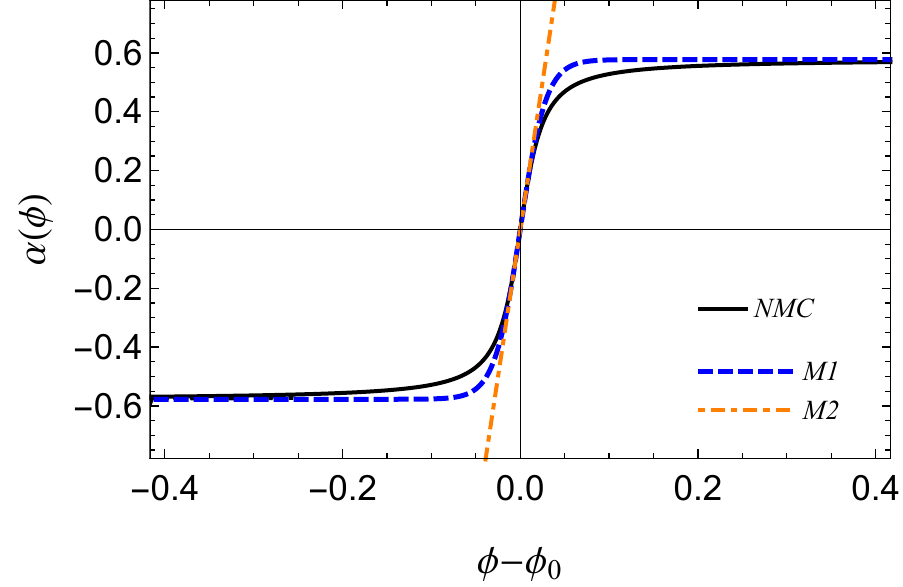}
\caption{The coupling function (\ref{eq:alpha}) as a function of the scalar field in the Einstein frame is shown in (solid) black, for $\xi = 10$. The blue (dashed) and orange (dashed-dotted) lines represent the coupling functions (\ref{eq:alpha1}) and (\ref{eq:alpha2}), respectively, with $\beta = 20$. M1 is seen to reproduce the qualitative features of the coupling function (\ref{eq:alpha}). For $\phi \approx \phi_0$, all curves overlap.}
\label{fig:NMC}
\end{figure}

Equation (\ref{eq:dPhi}) can be integrated to obtain $\phi(\Phi)$ and the inverse relation $\Phi(\phi)$. Note that, as $\Phi \to 0$, $\phi \to \phi_0 + 2\sqrt{\pi} \Phi + \mathcal{O}(\Phi^3)$, where $\phi_0$ is an integration constant. 
Ultimately, from Eqs.~(\ref{eq:dPhi}) and (\ref{eq:a_NMC}) we get the coupling function
\begin{equation}\label{eq:alpha}
\alpha(\phi) = \frac{4\sqrt{\pi} \xi \Phi(\phi)}{\sqrt{1-8\pi \xi (1-6\xi )\Phi(\phi)^2}},
\end{equation}
which is plotted in Fig.~\ref{fig:NMC}. Note that $\alpha(\phi_0) = 0$, $\alpha'(\phi_0) = 2\xi$, $\alpha''(\phi_0)=0$, $\alpha^{(3)}(\phi_0)=8(1-12\xi)\xi^2$ and so on.
Therefore, upon identifying $\xi = \beta/2$, the coupling functions (\ref{eq:alpha1}), (\ref{eq:alpha2}) and (\ref{eq:alpha}) all agree to linear order in $\phi-\phi_0$.
Moreover, M1 reproduces the qualitative behavior of the coupling function (\ref{eq:alpha}), and shares the distinctive feature that $\lim_{\phi\to\pm \infty} \alpha(\phi) = \pm 1/\sqrt{3}$ is finite, which will be relevant for the interpretation of our results.

\section{Relevant previous results}
\label{sec:previous}
\subsection{(In)Stability of GR-like solutions}
\label{sec:stability_GR}

Generically, the system of equations (\ref{eq:dm})-(\ref{eq:dp}) will admit at least one equilibrium solution, which is close to a solution in pure GR in the sense that the ratio of its scalar charge to the stellar mass, $\omega/M$, is of the order of $\alpha_0$. In particular, if $\alpha_0 = 0$, this solution consists of a scalar field everywhere equal to its cosmological value $\phi_0$, and metric and matter variables satisfying the usual Einstein-Euler equations. We will call these {\it GR solutions}. 

The stability of these GR solutions under linear scalar field perturbations was first analyzed in Ref.~\cite{Harada1997}, and it was shown that they may possess unstable scalar modes for a certain range of stellar compactnesses and coupling constants. Indeed, if we consider $\phi = \phi_0 + \delta \phi$ and treat $\delta \phi$ as a small perturbation, then Eqs.~(\ref{eq:G_eq}) and (\ref{eq:phi_eq}) imply, to first order,
\begin{equation} \label{eq:delta_phi}
\nabla_\mu \nabla^\mu \delta \phi = -4\pi \beta_0 \tilde{T} \delta \phi, 
\end{equation}
where the metric and matter variables are solutions to the background GR equations\footnote{Note that, at the linear level, scalar field perturbations of GR solutions are decoupled from metric and fluid perturbations, and we can restrict attention to the scalar sector. For more general perturbative analyses, see e.g.~\cite{Sotani2005, Sotani2014}.}. Note that Eq.~(\ref{eq:delta_phi}) holds both for M1 and M2, or any model which agrees with them up to $O(\delta \phi)$ in the coupling function $\alpha(\phi)$.

\begin{figure}[t]
\includegraphics[width=8.5cm]{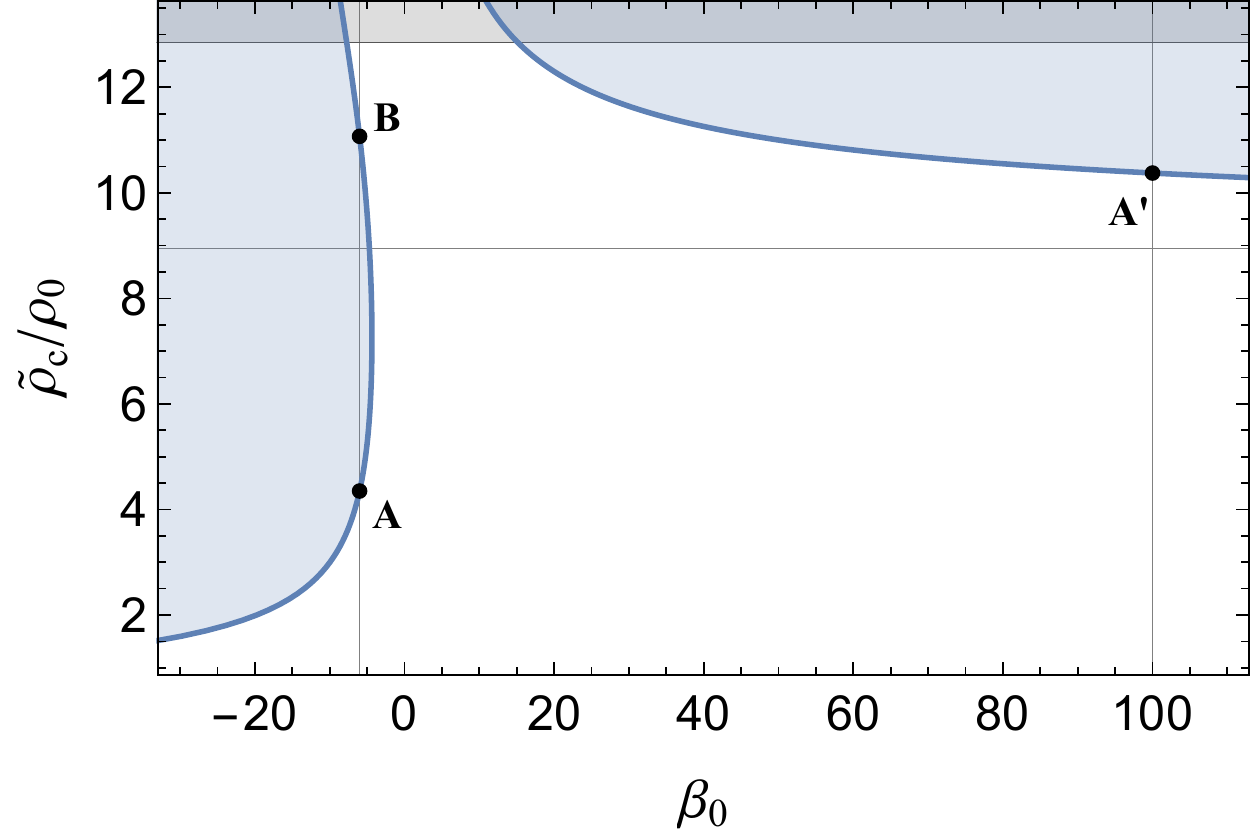}
\caption{Shaded blue regions correspond to values of $\beta_0$ and $\tilde{\rho}_c$ for which unstable modes of Eq.~(\ref{eq:delta_phi}) exist. The gray region on top indicates configurations which are hydrodynamically unstable according to GR. 
The condition $\tilde{T}(\tilde{\rho}_c) > 0$ is satisfied above the horizontal line at $\tilde{\rho}_c \simeq 8.94 \rho_0$, which also gives the asymptotic limit of the instability region to the right as $\beta_0 \to \infty$.
The vertical lines at $\beta_0=-6$ and $\beta_0=100$ highlight values that will be used in our analysis.}
\label{fig:inst1}
\end{figure}

The right-hand side of Eq.~(\ref{eq:delta_phi}) can be seen as containing an effective mass squared term, $m_\textrm{eff}^2 := -4\pi \beta_0 \tilde{T}$, and the fact that this is negative when $\beta_0 \tilde{T} > 0$ suggests the existence of a tachyonic-like instability. 
In particular, for $\beta_0>0$, a necessary condition for the existence of unstable modes of Eq.~(\ref{eq:delta_phi}) is that the trace of the energy momentum-tensor of matter fields, $\tilde{T} = 3\tilde{p}-\tilde{\epsilon}$, be positive in some region inside the star. A crucial property of the EoS we adopted, which is shared by several (but not all) realistic EoS, is that this condition is satisfied by some neutron stars which are hydrodynamically stable according to GR. This opens up a window where predictions from GR and STTs with $\beta_0>0$ differ, and can potentially be tested. 

\begin{figure}[b]
\includegraphics[width=8.5cm]{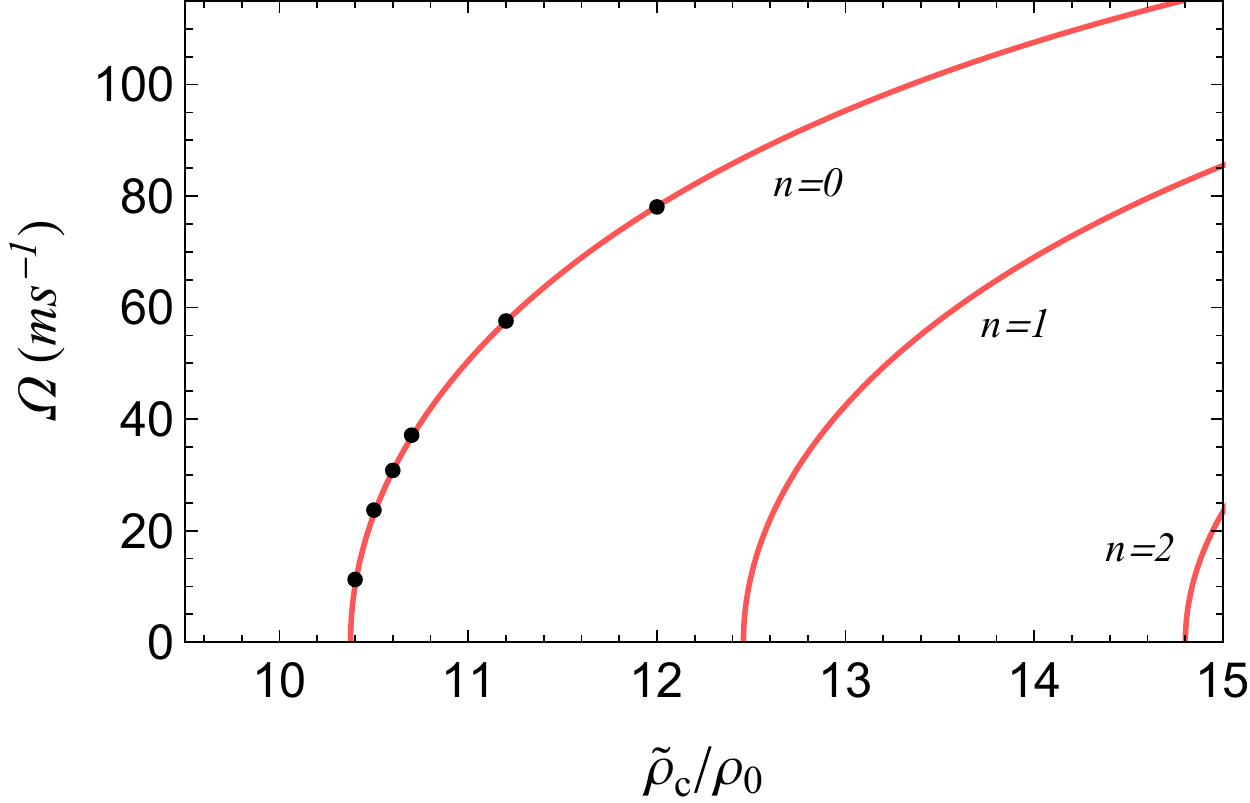}
\caption{Inverse time scale of the instability as a function of the star's central density, for $\beta_0=100$. A first unstable mode appears for stars with central density $\tilde{\rho}_c \gtrsim 10.38 \rho_0$. For $\tilde{\rho}_c \gtrsim 12.46 \rho_0$ and $\tilde{\rho}_c \gtrsim 14.80 \rho_0$, respectively, a second and a third scalar mode become unstable. The unstable spherical modes can be labeled by an overtone number $n$ which also measures the number of nodes in the radial profile of the scalar field and which is represented in the figure. Black dots indicate the rates of growth inferred from the numerical evolution of initial data consisting of unstable GR solutions with central densities $\tilde{\rho}_c/\rho_0 \in \{10.4,10.5,10.6,10.7,11.2,12\}$, as described in the main text.
}
\label{fig:inst2}
\end{figure}

A numerical search for solutions of Eq.~(\ref{eq:delta_phi}) of the form $\delta \phi = e^{\Omega t} f(r) $ in a spherically symmetric and static background, reveals that these solutions exist in the regions in parameter space shown in Fig.~\ref{fig:inst1}. The inverse time scale $\Omega$ of the instability is shown in Fig.~\ref{fig:inst2} as a function of the star's central density for $\beta_0=100$. In this case, a first unstable mode develops for stars with central density $\tilde{\rho}_c \gtrsim 10.38 \rho_0$, a second unstable mode appears when $\tilde{\rho}_c \gtrsim 12.46 \rho_0$, and a third one, when $\tilde{\rho}_c \gtrsim 14.80 \rho_0$. Indeed, as we increase the central density, we find a hierarchy of additional modes becoming unstable. These unstable modes can be labeled by a radial overtone number $n$, as indicated in Fig.~\ref{fig:inst2}, which also measures the number of nodes of the corresponding function $f(r)$. The $n=0$ mode is the fastest growing.
We remark that the existence of more than one unstable mode is not a particular feature of the $\beta_0>0$ case and can also occur for sufficiently negative values of $\beta_0$. 

The rate of growth of the unstable modes predicted by the linear stability analysis can be directly compared with the outcome of full nonlinear numerical evolutions. Therefore, as a test of our code, we set initial data consisting of GR solutions with central densities ranging from $10.2 \rho_0$ to $12 \rho_0$, and let them evolve according to the full nonlinear equations in either M1 or M2. We verify that configurations predicted to be stable simply oscillate around the equilibrium solution, and that those predicted to be unstable (i.e., with $\tilde{\rho}_c \gtrsim 10.38 \rho_0$) indeed undergo an initial phase of exponential growth of the scalar field. This is due to the fact that the initial field profile is not perfectly constant, but has fluctuations dictated by the size of the numerical errors, which are enough to trigger the instability. 
The black points in Fig.~\ref{fig:inst2} show the rate of exponential growth inferred from the numerical data, which is computed from the time it takes for $|\phi_c -\phi_0|$ to grow from $10^{-14}$ to $10^{-3}$. In this regime of small field amplitudes, the linear analysis is expected to hold, and indeed we find agreement between its predictions and the nonlinear evolution to a good extent, as shown in Fig.~\ref{fig:inst2} (see the Appendix for convergence details).

We refer to \cite{Mendes2015,Harada1997} for more details on the instability of GR solutions under scalar field perturbations, and to \cite{Lima2010,Lima2010b,Mendes2014} for an interesting quantum counterpart of this instability.

\subsection{The $\beta_0<0$ case: Spontaneous scalarization}
\label{sec:betaminus}

If $\beta_0$ is sufficiently negative ($\beta_0 \lesssim -4.35$ for our EoS), there is a range of central densities for which a GR solution with constant scalar field has unstable scalar modes (cf. Fig.~\ref{fig:inst1}). As we will review below, the onset of the instability of the GR solution is accompanied by the appearance of two additional stable equilibrium solutions, characterized by a nontrivial scalar field profile and scalar charge of the order of the stellar mass. Therefore, for a fixed value of $\beta_0<0$, if the stellar compactness increases above a critical threshold, the star undergoes a phase transition to a scalarized configuration. This spontaneous scalarization effect in STTs with $\beta_0<0$ has been extensively studied in the literature, both for the case of nonminimally coupled scalar fields, to which M1 is a rough approximation and, more prolifically, for M2 (see e.g. \cite{Berti2015a} for a literature survey). In this section we briefly dwell on the case of $\beta<0$ in Eqs.~(\ref{eq:alpha1}) and (\ref{eq:alpha2}) in order to compare it with our main results in Sec.~\ref{sec:betaplus}.

\begin{figure}[tb]
\includegraphics[width=8.5cm]{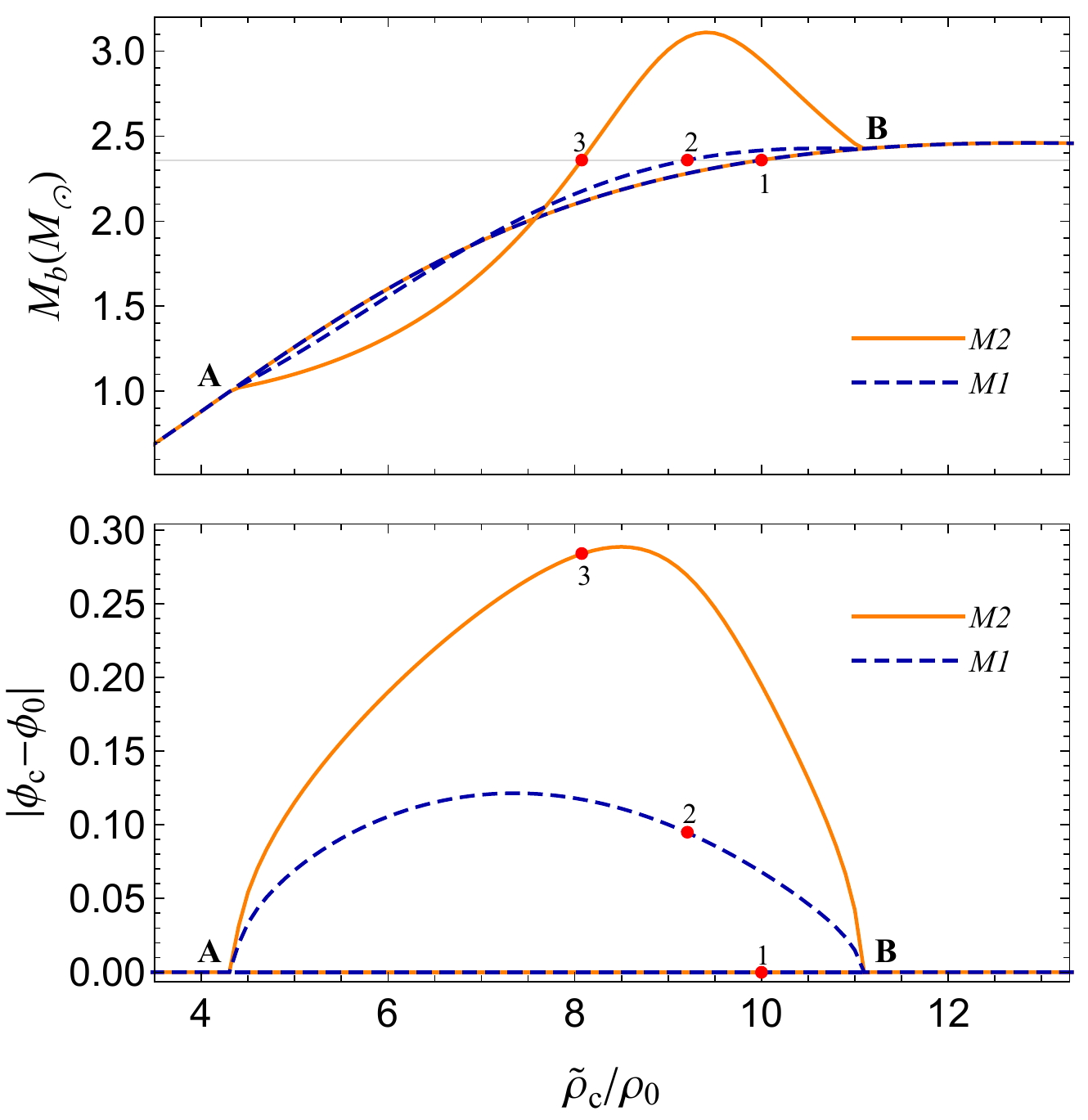}
\caption{Baryonic mass and central scalar field as functions of the central rest-mass density for sequences of equilibrium solutions in M1 (dashed blue) and M2 (solid orange) with $\beta=-6$.
For $\tilde{\rho}_c \lesssim 4.38 \rho_0$ and $\tilde{\rho}_c \gtrsim 11.15 \rho_0$, only the GR branch of solutions exists, which is the same in both models. For $4.38 \lesssim \tilde{\rho}_c/\rho_0 \lesssim 11.15$ (between points {\bf A} and {\bf B}; see also Fig.~\ref{fig:inst1}), two extra (degenerate) branches appear, characterized by a nontrivial scalar field profile and nonzero scalar charge. We highlight solutions with baryonic mass $M_b = 2.3594 M_\odot$, described in Table \ref{table:initialdata1}.}
\label{fig:betaminus}
\end{figure}

Before studying the development of the instability discussed above from a dynamical perspective, it is instructive to review some properties of static solutions, in order to determine the possible equilibrium configurations to which the unstable system could in principle settle.
Fixing a value of $\beta$, say, $\beta=-6$, we construct equilibrium solutions of Eqs.~(\ref{eq:dm})-(\ref{eq:dp}) in Models 1 and 2 for a range of central densities. The baryonic mass and the central value of the scalar field of the solutions are shown as a function of the central density in Fig.~\ref{fig:betaminus}.
The overall picture is the same for both models: for low central densities, only one equilibrium solution exists, the GR solution discussed above. Then, precisely at the critical density above which this branch of solutions becomes unstable under scalar field perturbations (point \textbf{A} in Figs.~\ref{fig:inst1} and \ref{fig:betaminus}), another branch of solutions develops\footnote{Actually, two new branches develop, as discussed, e.g., in Ref.~\cite{Harada1998}; however, for $\alpha_0=0$, they are trivially related by the transformation $\phi \to -\phi + 2 \phi_0$, and have the same global properties, such as mass or (absolute value of the) scalar charge. If $\alpha_0 \neq 0$, these solutions are no longer degenerate, but the property that $\omega/M\sim 1$ still holds for both.}, characterized by a nontrivial profile of the scalar field. We will call this the \textit{scalarized branch}.
These solutions can be parametrized by their scalar charge or the difference between the central and the asymptotic value of the scalar field: from Fig.~\ref{fig:betaminus} we see that, as we move along the scalarized branch, $|\phi_c - \phi_0|$ increases from zero to a maximum value, and then decreases to zero as this branch finally merges with the GR one, at the density above which the latter is no longer unstable under scalar field perturbations (point \textbf{B} in Figs.~\ref{fig:inst1} and \ref{fig:betaminus}).

\begin{table}[b]
\begin{tabular}{ | l | c | c | c | c | c | c | }
	\hline                       
	Solution & $\tilde{\rho}_c/\rho_0$ & $M_b [M_\odot]$ & $M [M_\odot]$ & $M/R_s$ & $|\phi_c\!-\!\phi_0|$ & $| \omega | [M_\odot]$  \\
	\hline
	1*       & 10.0 & 2.3594 & 1.9650 & 0.287 & 0 & 0 \\
	2 (M1) & 9.2061 & 2.3594 & 1.9641 & 0.273 & 0.095 & 0.30 \\
	3 (M2) & 8.0747 & 2.3594 & 1.9459 & 0.209 & 0.284 & 1.23 \\
	\hline  
\end{tabular}
\caption{Properties of some equilibrium solutions in STTs with $\beta=-6$, namely, central density, baryonic mass, total mass, compactness, central scalar field, and scalar charge. The solution marked with a star is used as initial data for the numerical simulation shown in Fig.~\ref{fig:betaminusEvol}.}
\label{table:initialdata1}
\end{table}

\begin{figure}[b]
\includegraphics[width=8.5cm]{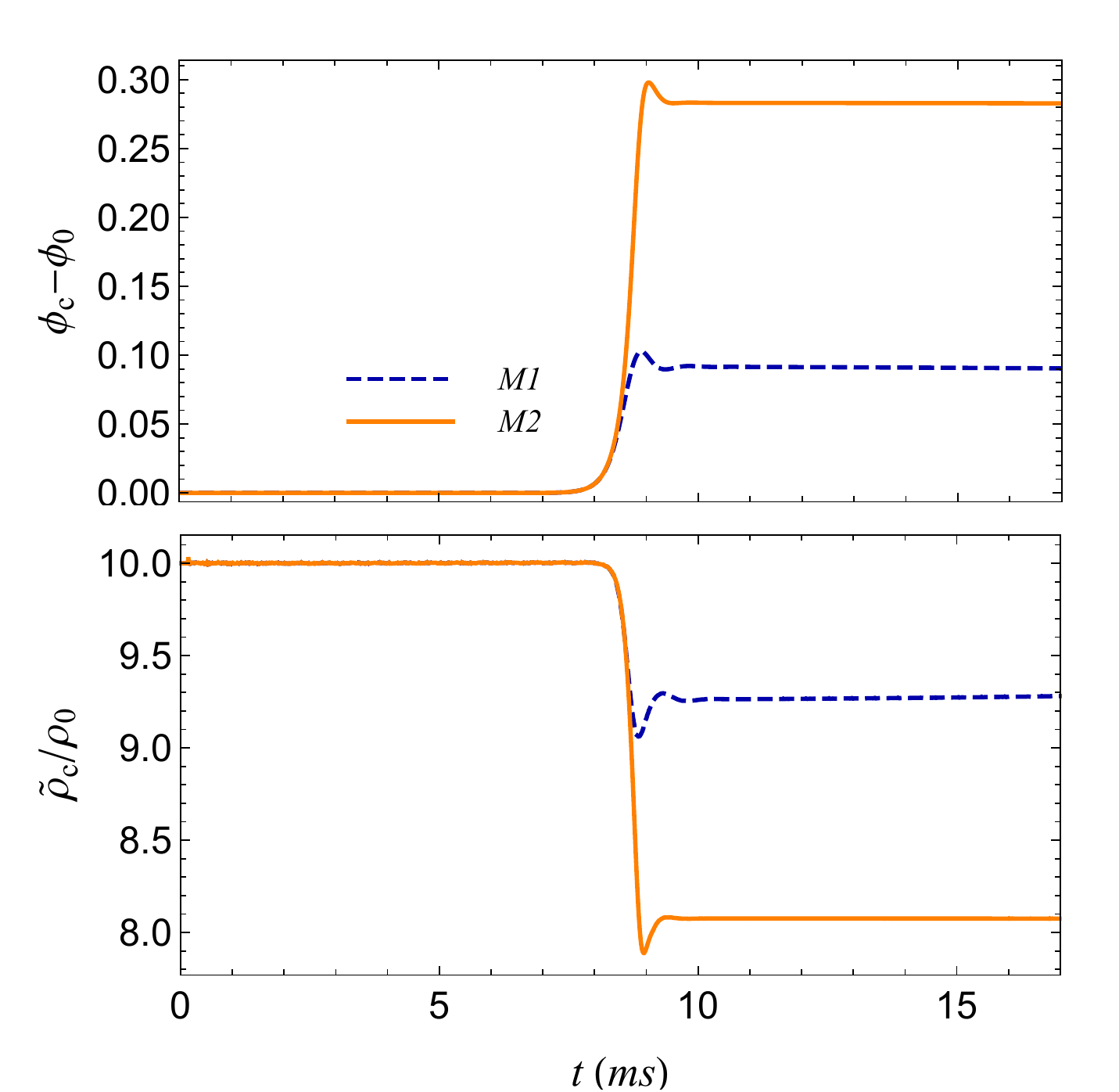}
\caption{Evolution of the central value of the scalar field and of the star's central density for the unstable initial data labeled 1 in Table~\ref{table:initialdata1}, within Models 1 and 2.}
\label{fig:betaminusEvol}
\end{figure}

Thermodynamic arguments indicate that solutions in the scalarized branch are stable up to the turning point in the $(\tilde{\rho}_c,M_b)$ diagram \cite{Harada1998}. Moreover, for a given baryonic mass, a scalarized configuration always has a smaller total mass than the corresponding GR solution and, in this sense, the former is energetically favored over the latter. This is illustrated in Table \ref{table:initialdata1} for the (red) points highlighted in Fig.~\ref{fig:betaminus}. The fact that scalarized configurations are stable and energetically favored over the unstable GR solutions suggests that they are plausible end states of the evolution.
Indeed, full nonlinear numerical simulations have confirmed that this is the case. The dynamical transition from an unstable GR configuration to a scalarized one has been studied numerically both for M2 \cite{Novak1998} and for nonminimally coupled scalar fields \cite{Alcubierre2010,Ruiz2012}. 
In Fig.~\ref{fig:betaminusEvol} we show illustrative results from our numerical simulations. The initial data is chosen to be a GR solution with $\tilde{\rho}_c = 10 \rho_0$, whose properties are shown in the first row of Table \ref{table:initialdata1}. The magnitude of the initial field fluctuation is simply determined by round-off errors, which are typically of the order $10^{-16}$. This is enough, however, to trigger the instability described in Sec.~\ref{sec:stability_GR}. The evolution in time of the central value of the scalar field and the central rest-mass density is shown in Fig.~\ref{fig:betaminusEvol}. After an initial phase of exponential growth, the field settles down to a scalarized equilibrium configuration, which is consistent with solutions 2 and 3 in Table \ref{table:initialdata1}, for Models 1 and 2, respectively. It is worth emphasizing that although quantitative aspects of the final configuration, such as total mass or scalar charge, depend on the details of the coupling function $\alpha(\phi)$, the fact that the final state is a scalarized configuration is generic for theories with the same value of $\beta_0 = \alpha'(\phi_0)$. As we will see in the next section, this is no longer true when $\beta_0>0$.

\section{The $\beta_0>0$ case: scalarization {\it vs.} gravitational collapse}
\label{sec:betaplus}

In this section we present our main results regarding the nonlinear development of the instability discussed in Sec.~\ref{sec:stability_GR} for theories with $\beta_0>0$. Given the different outcomes observed for Models 1 and 2, we present the results for each of them separately below.

\subsection{Model 1}
\label{sec:M1}

In Ref.~\cite{Mendes2015}, scalarized equilibrium solutions for neutron stars within the model of nonminimally coupled scalar fields [cf.~Eq.~(\ref{eq:S_NMC})] were already constructed in the $\beta_0>0$ regime (see also Ref.~\cite{Pani2011}).
Here we explore more thoroughly the properties of these solutions within Model 1, together with numerical simulations of the stellar evolution in order to determine the final state of unstable configurations. 

\begin{figure}[tbh]
\includegraphics[width=8.5cm]{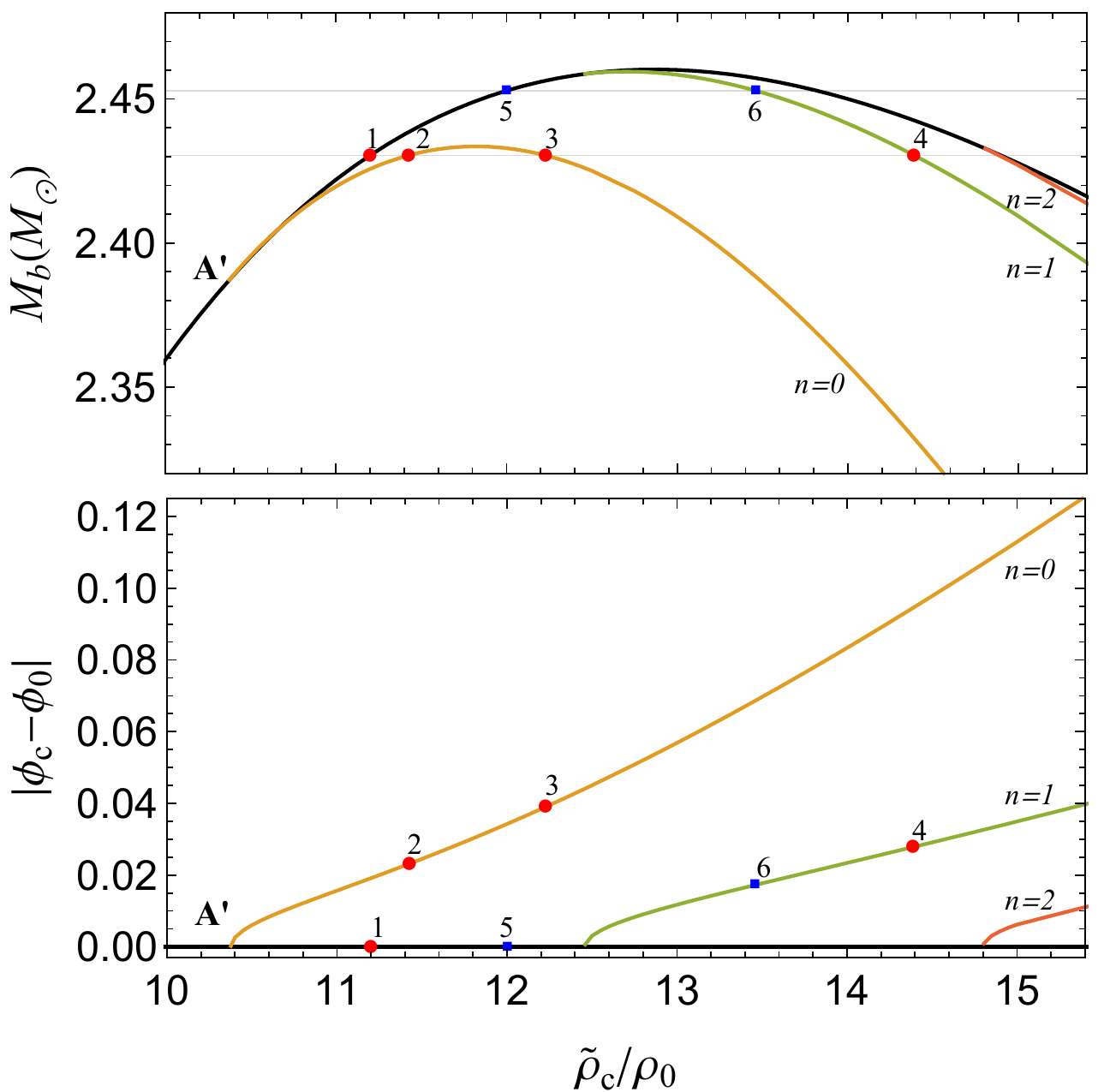}
\caption{Baryonic mass and central scalar field as functions of the central rest-mass density for a sequence of equilibrium solutions in M1, with $\beta = 100$. For $\tilde{\rho}_c \lesssim 10.38 \rho_0$ (point $\textbf{A}'$ in Fig.~\ref{fig:inst1}), only the GR solution exists. Above this value, a branch of scalarized solutions develops, along which the absolute value of the scalar charge increases monotonically. As the central density increases, we see more branches of scalarized solutions developing from the unstable GR branch. An overtone number $n$ is assigned to each of them. We highlight some solutions with baryonic mass $M_b=2.4304 M_\odot$ (points 1 to 4, in red) and $M_b=2.4529 M_\odot$ (points 5 and 6, in blue), described in Table \ref{table:initialdata2}.}
\label{fig:betaplus1}
\end{figure}

We begin by constructing equilibrium solutions of Eqs.~(\ref{eq:dm})-(\ref{eq:dp}) in M1 for a range of central densities. Figure \ref{fig:betaplus1} shows the baryonic mass and central scalar field of these solutions as a function of the star's central density, in the case of $\beta = 100$. Again, for low central densities there is only the trivial, GR equilibrium solution; then, at a critical density corresponding to the onset of the instability in the GR branch (point $\textbf{A}'$ in Figs.~\ref{fig:inst1} and \ref{fig:betaplus1}), a scalarized branch develops. 
Differently from the $\beta<0$ case, the scalarized branch does not rejoin the GR one at higher densities, and the (absolute value of the) scalar charge grows monotonically along it. 

Notice, from Fig.~\ref{fig:betaplus1}, the existence of a second and third branches of scalarized solutions that detach from the GR branch at higher densities. In fact, as we increase the central density for a fixed value of $\beta>0$, we find a hierarchy of such solutions, each new branch characterized by a scalar field profile with a higher number of nodes. The appearance of these ``excited'' solutions (observed also in Ref.~\cite{Pani2011}) has an equivalent in the linear stability analysis as well, as discussed in Sec.~\ref{sec:stability_GR}. Indeed, at each critical density where a scalarized branch starts, we find a new scalar mode of the GR solution becoming unstable (see Fig.~\ref{fig:inst2}). We will refer to each of these scalarized branches by an assigned radial overtone number $n$, corresponding to the number of nodes in the radial profile of the scalar field. 

\begin{figure}[thb]
\includegraphics[width=8.5cm]{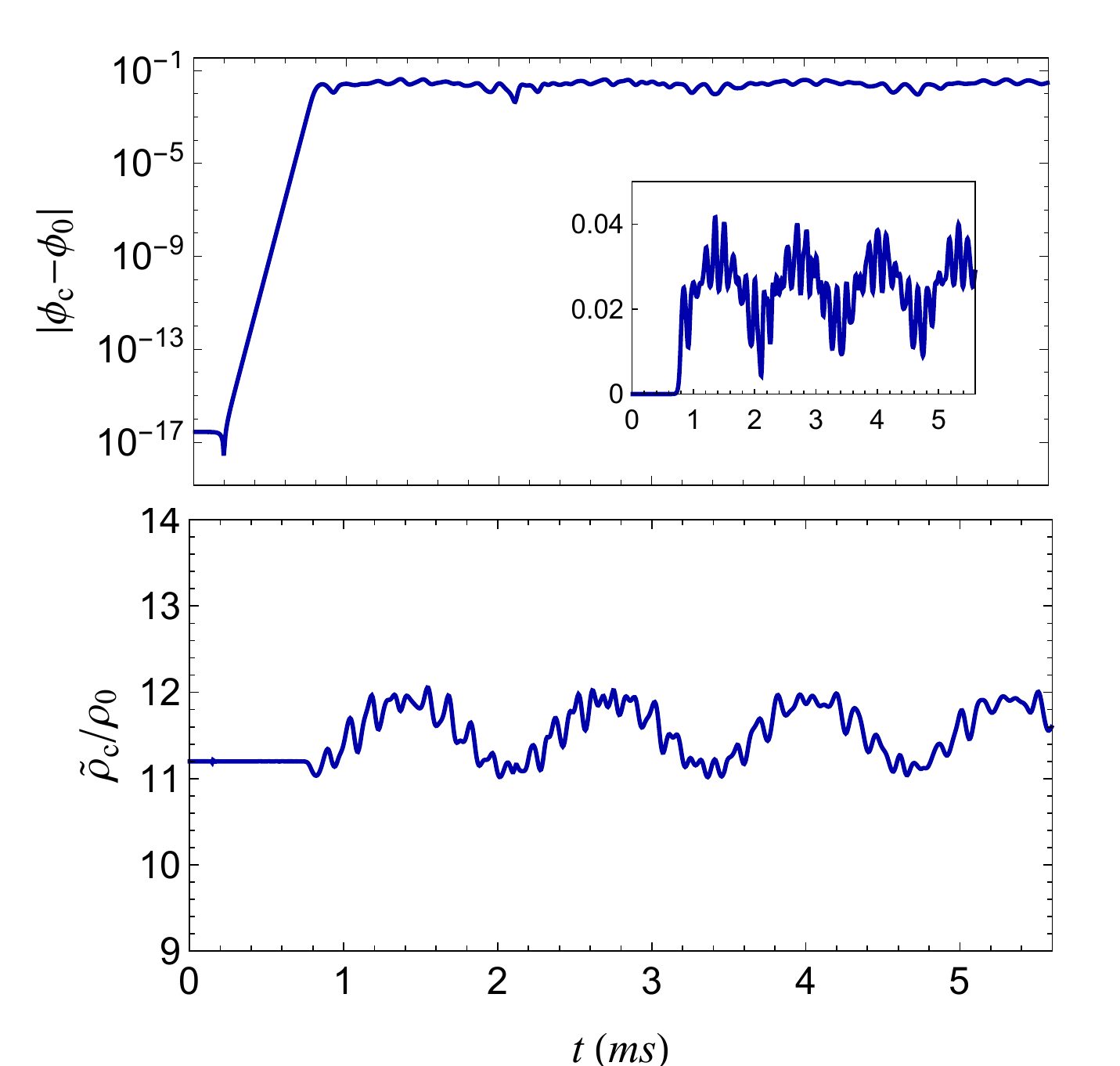}
\caption{Evolution of the central value of the scalar field and of the star's central density for an initial unstable equilibrium solution (solution 1 in Table~\ref{table:initialdata2}).}
\label{fig:betaplusEvol1}
\end{figure}

Our numerical experiments indicate that solutions in the $n=0$ scalarized branch are stable up to the turning point in the $(\tilde{\rho}_c,M_b)$ diagram of Fig.~\ref{fig:betaplus1}, which is consistent with thermodynamic expectations. 
Moreover, we find that stable scalarized configurations have lower total mass than unstable GR solutions with the same baryonic mass, as can be seen by comparing, e.g., solutions 1 and 2 in Table \ref{table:initialdata2}. 
The expectation that unstable GR solutions may settle to these scalarized configurations is corroborated by our numerical simulations. In particular, in Fig.~\ref{fig:betaplusEvol1} we show the evolution of the initial data labeled 1 in Table \ref{table:initialdata2}, which consists of a GR solution with $\tilde{\rho}_c = 11.2 \rho_0$. The scalar field, whose central value is displayed in the upper panel of Fig.~\ref{fig:betaplusEvol1}, goes through a phase of exponential growth, starting from an initial value dictated by the size of round-off errors. The rate of exponential growth inferred from the numerical data is consistent with the prediction from the linear theory, as was shown in Fig.~\ref{fig:inst2}. After the field reaches a value of the order of $10^{-2}$, its growth is quenched and it oscillates around a value consistent with solution 2 in Table \ref{table:initialdata2}. Since the final central value of the scalar field is much smaller than the example shown in Sec.~\ref{sec:betaminus}, the relative size of the oscillations in the present case is much larger than what is seen in Fig.~\ref{fig:betaminusEvol}.

\begin{table}[b]
\begin{tabular}{ | l | c | c | c | c | c |}
	\hline                       
	Solution & $\tilde{\rho}_c/\rho_0$ & $M_b [M_\odot]$ & $M [M_\odot]$ & $M/R_s$ & $|\phi_c-\phi_0|$ \\
	\hline
	\, 1* & 11.20 & 2.4304 & 2.01058 & 0.302 & 0 \\
	\, 2 & 11.4251 & 2.4304 & 2.01053 & 0.304 & 0.023159  \\
	\, 3 & 12.2279 & 2.4304 & 2.01056 & 0.311 & 0.039036 \\
	\, 4 & 14.3890 & 2.4304 & 2.01132 & 0.325 & 0.027827  \\
	\, 5* & 12.0 & 2.4529 & 2.02461 & 0.309 & 0  \\
    \, 6 & 13.4600 & 2.4529 & 2.02469 & 0.320 & 0.017240 \\
	\hline  
\end{tabular}
\caption{Properties of some equilibrium solutions with $M_b = 2.4304 M_\odot$ (solutions 1 to 4) and $M_b = 2.4529 M_\odot$ (solutions 5 and 6) in M1 with $\beta=100$.
Solutions marked with stars are used as initial data for numerical simulations.
}
\label{table:initialdata2}
\end{table}

\begin{figure}[b]
\includegraphics[width=8.5cm]{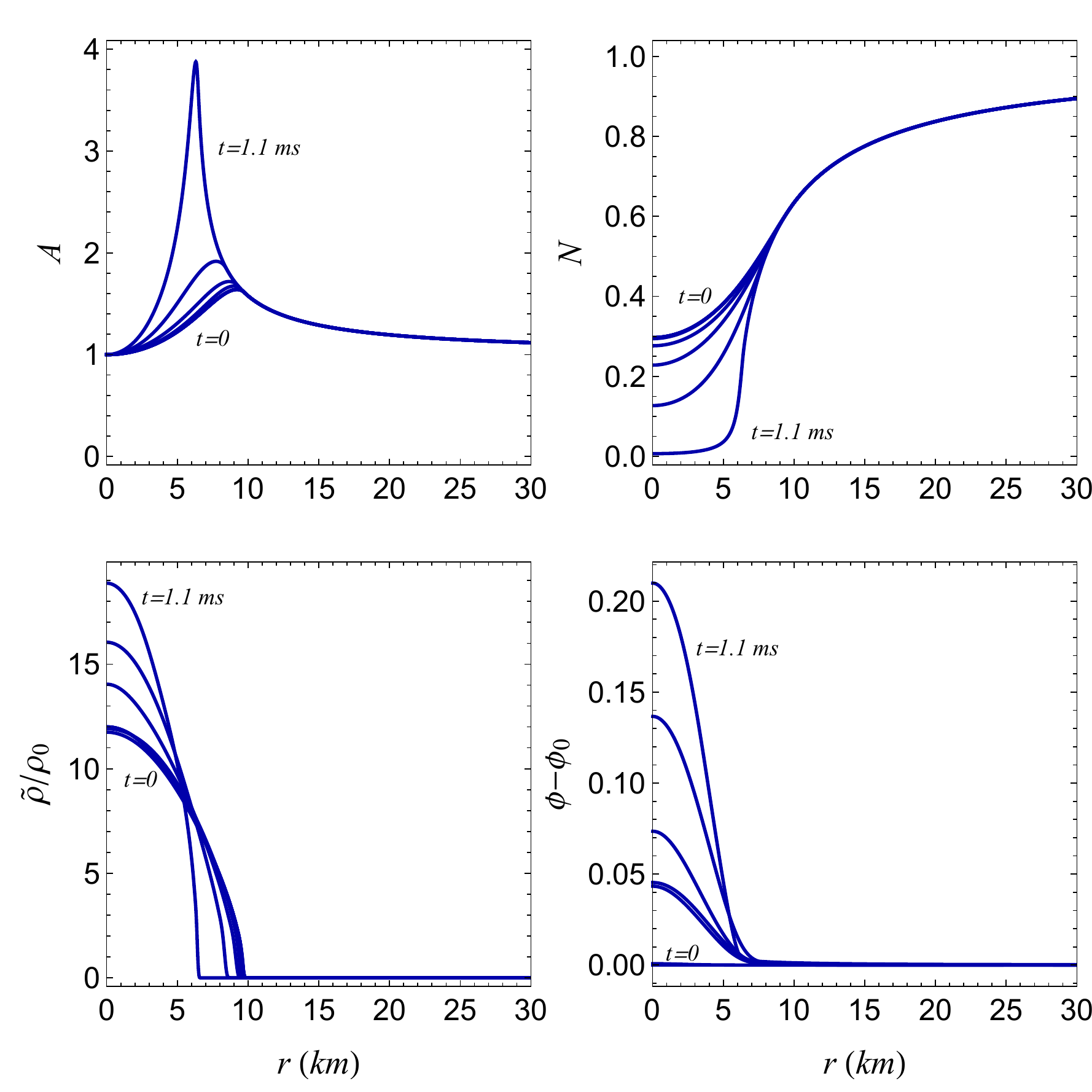}
\caption{Snapshots of the time evolution of the metric functions $A$ and $N$, of the rest-mass density $\tilde{\rho}$ and the scalar field $\phi$ for an initial unstable equilibrium solution (solution 5 in Table~\ref{table:initialdata2}).}
\label{fig:betaplusEvol2}
\end{figure}

An important difference from the $\beta<0$ case discussed in Sec.~\ref{sec:betaminus} is that the maximum baryonic mass of the $n=0$ scalarized branch is now lower than that of the GR branch. This is consistent with the interpretation that, when $\beta>0$ in M1, the effective gravitational coupling $G_\textrm{eff}=G a(\phi)^2$ is larger in regions with large field amplitudes, and not-as-massive scalarized stars are supported without undergoing gravitational collapse.
Also, this immediately indicates that there is a set of unstable GR solutions that cannot evolve to stable, scalarized configurations while conserving baryonic mass, and must meet a different fate. In Fig.~\ref{fig:betaplusEvol2} we show snapshots of the numerical evolution of an unstable GR solution with $\tilde{\rho}_c = 12 \rho_0$ (solution 5 of Table \ref{table:initialdata2}). We see from Fig.~\ref{fig:betaplus1} that there are no stable scalarized solutions with the same baryonic mass as this one. In accordance with this, our numerical simulations show that this configuration collapses to a black hole, in a time scale dictated by the growth rate of the unstable mode. 

It is opportune to mention the criterion we adopt to decide when a star has collapsed to a black hole, which is based on the behavior of the metric functions.
Our slicing choice---polar slicing---has the good property of avoiding the physical singularity at $r=0$; however, it does not capture the formation of an apparent horizon, which would be a direct evidence of black hole formation. Instead, as we approach the formation of an event horizon, the lapse function $N$ exponentially shrinks to zero.
This causes a {\it slice stretching} effect, which reflects on the rapid development of large gradients in the radial metric component $A$, ultimately leading the code to crash~\cite{Baumgarte2010}. Nonetheless, the characteristic behavior of the metric components in this gauge as a star collapses is enough for our purpose of determining the fate of initially unstable configurations. An example can be seen in the upper panels of Fig.~\ref{fig:betaplusEvol2}.

Finally, it is interesting to notice that the $n=1$ scalarized branch in Fig.~\ref{fig:betaplus1} also seems to have a small stable portion, in which the solutions have a marginally smaller total mass than a solution in the GR branch with the same baryonic mass. However, due to the proximity to the turning point, we were not able to determine from numerical simulations whether these solutions could be the end state of unstable GR configurations.

Our results reveal that the end state of the instability discussed in Sec.~\ref{sec:stability_GR} may vary from spontaneous scalarization to gravitational collapse in Model 1. In particular, they show that spontaneous scalarization is not unique to theories with $\beta_0<0$ but can also happen in theories with $\beta_0>0$. We discuss some implications of these results in Sec.~\ref{sec:discussion}.

\subsection{Model 2}
\label{sec:M2}

In Ref.~\cite{Palenzuela2016}, the evolution of unstable GR-like configurations was investigated within a slight variant of Model 2, where a nonzero but small value of $\alpha_0$ was implicitly included. The outcome of the simulations was shown to be collapse to a black hole, although some long-lived oscillating configurations were also reported. Here, we present the results of similar numerical analyses. Our setup slightly differs from the one in Ref.~\cite{Palenzuela2016}, in particular because the initial data chosen in that work, which consists of a solution to the Einstein-Euler equations and a constant scalar field profile, is not an actual equilibrium solution when $\alpha_0 \neq 0$, and this introduces spurious dynamics in the evolution. Notwithstanding, our numerical simulations mostly agree with their results: unstable GR solutions are seen to undergo gravitational collapse.

\begin{figure}[b]
\centering
\includegraphics[width=8.5cm]{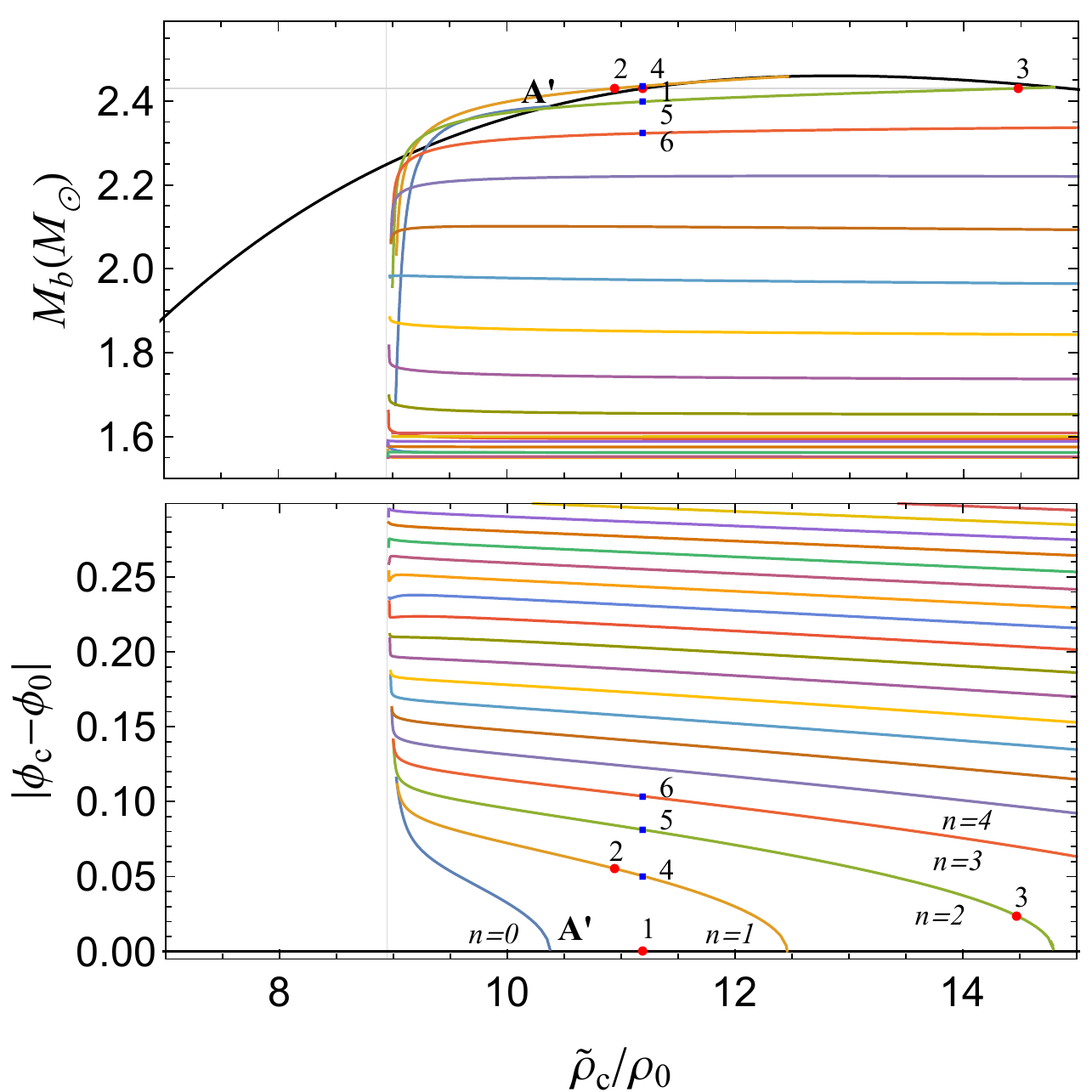}
\caption{Baryonic mass and central scalar field as functions of the central rest-mass density for a sequence of equilibrium solutions in M2, with $\beta = 100$. We only display the scalarized solutions found with $|\phi_c-\phi_0| < 0.3$, but additional solutions exist beyond this point. Each scalarized branch is characterized by the number of nodes in the scalar field profile, as indicated in the lower panel for the first five branches.
Note that solutions with nonzero scalar charge are found for $\tilde{\rho}_c < 10.38 \rho_0$ (point $\textbf{A}'$; cf.~Fig.~\ref{fig:inst1}), i.e., before the onset of instability in the GR branch.
A vertical line at $\tilde{\rho}_c = 8.94 \rho_0$ highlights the central density above which the trace of the energy-momentum tensor is positive at the stellar center. 
We also highlight some solutions with baryonic mass $M_b=2.4304 M_\odot$ (points 1 to 3, in red) and additional solutions with  $\tilde{\rho}_c = 11.20 \rho_0$ (points 4 to 6, in blue), which are described in Table \ref{table:initialdata3}.
}\label{fig:betaplus2}
\end{figure}

We can gain additional understanding on this result from an energy balance analysis of static equilibrium configurations.
Indeed, scalarized equilibrium solutions in Model 2 display a startling different behavior from those in Model 1. 
Sequences of equilibrium configurations, obtained by solving Eqs.~(\ref{eq:dm})-(\ref{eq:dp}), are shown in Fig.~\ref{fig:betaplus2} for M2 with $\beta = 100$, although there is a similar behavior for any $\beta > 0$.
Unlike what we have seen so far, scalarized equilibrium solutions exist even for values of the central density for which the GR solution does not possess any unstable scalar mode. In fact, a hierarchy of solutions is found for $\tilde{\rho}_c \gtrsim 8.94 \rho_0$, i.e., for central densities such that $\tilde{T}(\tilde{\rho}_c) > 0$.
Again, each scalarized branch is characterized by a number of nodes in the radial profile of the scalar field, and a radial overtone number $n$ is assigned accordingly.
Similarly to M1, these scalarized branches detach from the GR branch at values of $\tilde{\rho}_c$ for which some spherical scalar mode of the GR solution becomes unstable. This is clearly visible in Fig.~\ref{fig:betaplus2} for the scalarized branches with $n=0$, $n=1$, and $n=2$, but holds for higher overtones as well. However, differently from M1, these branches develop toward lower densities, approaching in different ways the critical central density $\tilde{\rho}_c \simeq 8.94 \rho_0$.

\begin{table}[t]
\begin{tabular}{ | l | c | c | c | c | c |}
	\hline                       
	Solution & $\tilde{\rho}_c/\rho_0$ & $M_b [M_\odot]$ & $M [M_\odot]$ & $M/R_s$ & $|\phi_c-\phi_0|$ \\
	\hline
	\, 1* & 11.20 & 2.4304 & 2.01058 & 0.302 & 0 \\
	\, 2 & 10.9493 & 2.4304 & 2.01087 & 0.312 & 0.055  \\
	\, 3 & 14.4841 & 2.4304 & 2.01142 & 0.327 & 0.023 \\
	\, 4* & 11.20 & 2.43611 & 2.01435 & 0.313 & 0.050  \\
	\, 5* & 11.20 & 2.39877 & 1.99273 & 0.325 & 0.081  \\
    \, 6* & 11.20 & 2.32321 & 1.94928 & 0.332 & 0.104 \\
	\hline
\end{tabular}
\caption{Properties of some equilibrium solutions with $M_b = 2.4304 M_\odot$ (solutions 1 to 3) and $\tilde{\rho}_c = 11.2 \rho_0$ (solutions 1 and 4 to 6) in M2 with $\beta=100$.
Solutions marked with stars are used as initial data for numerical simulations.}
\label{table:initialdata3}
\end{table}

As illustrated in Table \ref{table:initialdata3}, all scalarized solutions in this case have a higher total mass for a fixed baryonic mass than the GR one, which suggests they all are unstable, and likely unphysical. This is in agreement with the numerical results of Ref.~\cite{Palenzuela2016}, which exhibited no sign of spontaneous scalarization, and with our own numerical experiments, which we discuss below. 
Before that, however, let us try to understand why scalarized equilibrium solutions in M2 with $\beta>0$ are so different from the $\beta<0$ case and from M1 with $\beta>0$.
The crucial equations here are (\ref{eq:dphi}) and (\ref{eq:dp}), which we repeat below:
\begin{align*}
 &\phi'' = \frac{4\pi r A^4}{r-2m} \left[\alpha \overset{\mathbf{(I)}}{(\tilde{\epsilon} - 3\tilde{p})} + r (\tilde{\epsilon} - \tilde{p}) \phi' \right ] -\frac{2(r-m)}{r(r-2m)} \phi' \\
 &\tilde{p}' = -(\tilde{\epsilon} + \tilde{p}) \left[  \frac{4\pi r^2 A^4 \tilde{p}}{r-2m} + \frac{r}{2} \phi'^2 + \frac{m}{r(r-2m)} + \overset{\mathbf{(II)}}{\alpha \phi'} \right],
\end{align*}
with a prime denoting derivative with respect to $r$. 
Let us consider Model 2, where $\alpha(\phi) = \beta \phi$ (we set $\phi_0=0$ for definiteness), and assume, with no loss of generality, that $\phi_c>0$. Near $r=0$, for a sufficiently large value of $|\beta|$, term (I) dominates in the first equation. If $\beta \tilde{T}_c = \beta (3\tilde{p}_c -\tilde{\epsilon}_c) > 0$, then $\phi'' < 0$ and $\phi'<0$ near the origin. The difference between the $\beta<0$ and $\beta>0$ cases arises from the feedback between terms (I) and (II): if $\beta < 0$, term (II) is positive, and contributes to a more rapid decrease of the pressure, lessening the weight of (I); however, if $\beta>0$, term (II) is negative, and can increase (or, at least, delay the decay of) the pressure. (Note that the right-hand side of the second equation would be strictly negative if it was not for term (II).) This reinforces term (I), making $\phi'$ even more negative. This positive feedback continues until the other terms eventually start dominating.
This argument is consistent with our findings that some equilibrium solutions for M2 show a ``pathological'' behavior, in the sense that $\phi'$ can increase negatively by orders of magnitude in a small spatial region near the origin, and the pressure is not necessarily a monotonically decreasing function of $r$ \cite{MathematicaNotebook}. 
It is also consistent with the tentative interpretation in terms of a field-dependent effective coupling constant $G_\textrm{eff}=G a(\phi)^2$: when $\beta>0$, gravity is ``stronger'' in regions with large field amplitudes, and scalarized configurations need more pressure to be supported; however, since the scalar field is sourced by pressure, this induces an increase in the scalar field amplitude and a strengthening of gravity, leading to the positive feedback we described.
The different behavior displayed by M1 is likely related to the fact that $\alpha(\phi)$---which determines the coupling of the scalar field to matter---is bounded in this case (see Fig.~\ref{fig:NMC}), and $G_\text{eff}$ increases only polynomially with $\phi$ as $\phi \to \infty$, instead of exponentially as in M2.

\begin{figure}[h]
\includegraphics[width=8.5cm]{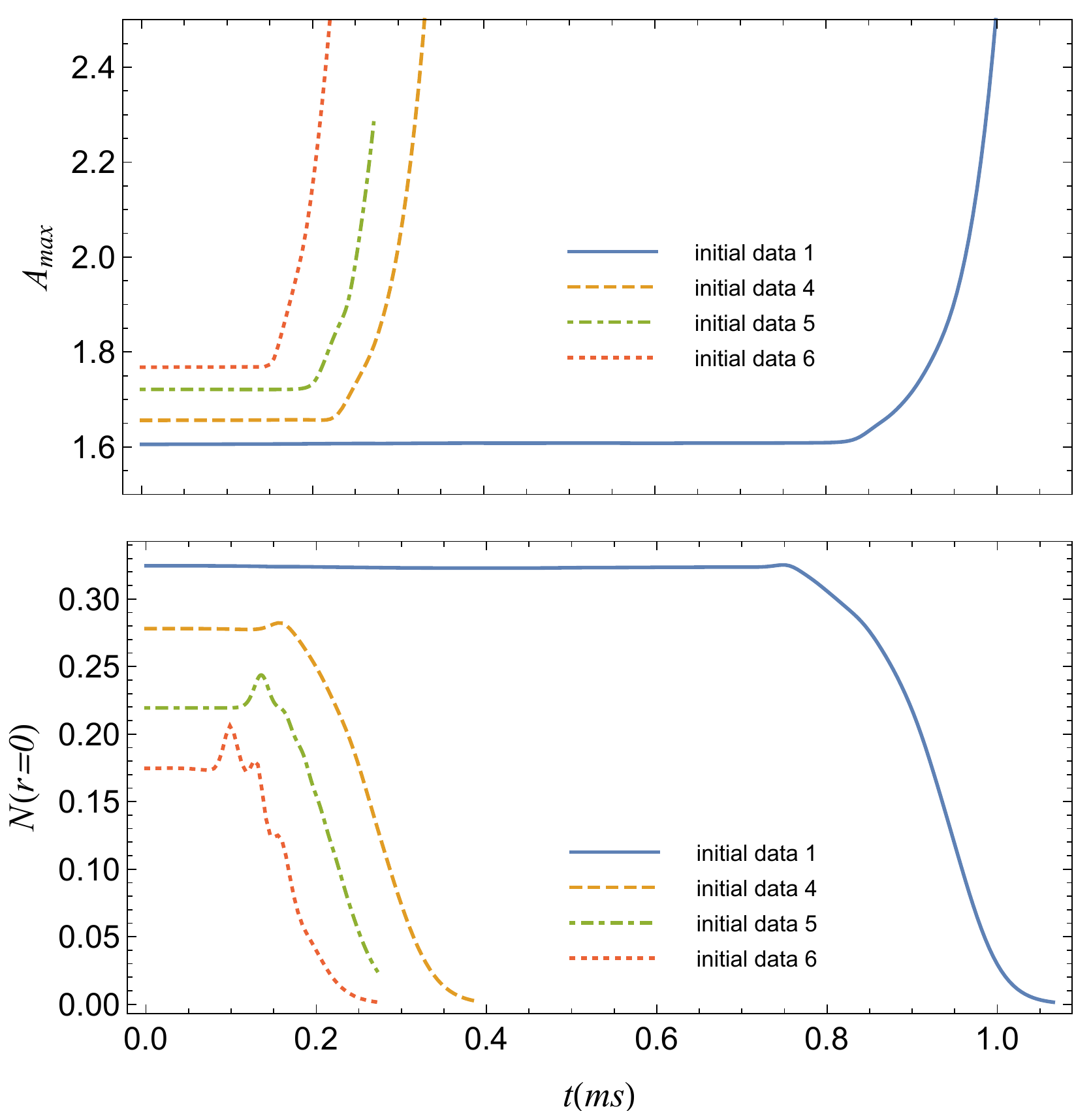}
\caption{Evolution of the maximum value of $A(r)$, and the central value of the lapse, $N(r=0)$, for some unstable equilibrium solutions (solution 1, 4, 5 and 6 in Table~\ref{table:initialdata3}).}
\label{fig:betaplusEvol3}
\end{figure}

From our numerical experiments with various initial data, consisting both of unstable GR solutions and scalarized solutions in any branch, we always find that the system evolves towards gravitational collapse.
Some examples are displayed in Fig.~\ref{fig:betaplusEvol3}, where we show the time evolution of initial data labeled 1, 4, 5 and 6 in Table \ref{table:initialdata3}. We plot the maximum value of $A(r)$, and the central value of the lapse, $N(r=0)$, as a function of time in these four cases. As we discussed in Sec.~\ref{sec:M1}, the first quantity diverges and the second one goes to zero as we approach the formation of an event horizon. This is seen to happen for all configurations, including the GR solution with $\tilde{\rho}_c = 11.2\rho_0$ which was shown to undergo spontaneous scalarization in Model 1 (see Fig.~\ref{fig:betaplusEvol1}). 
Notice, additionally, that the time scale of the collapse is shorter for higher overtone solutions.

\section{Discussion}
\label{sec:discussion}

Some highly compact neutron stars that are stable according to general relativity would be subject to an instability in some scalar-tensor theories of gravity with $\beta_0>0$.
In this work, we investigated the nonlinear development of such instability. Before discussing our results and their implications, let us briefly review the properties of the systems in which this instability arises. These are neutron stars in the core of which the pressure surpasses one third of the energy density, i.e. the trace of the fluid's energy-momentum tensor is positive at the stellar center. This condition, which is unachievable by free or weakly interacting particles, can be fulfilled by strongly interacting systems such as neutron stars. For different equations of state, this condition translates into a minimum required compactness; interestingly, however, this minimum compactness depends only weakly on the equation of state, being around $(M/R_s)_\textrm{min} \sim 0.265$ \cite{Mendes2015}. For a $2 M_\odot$ star this would require a radius $R_s \lesssim 11$km, which is consistent with predictions of several realistic equations of state and the current observational constraints \cite{Ozel2016}.

On the other hand, the lowest possible value of $\beta_0>0$ for which this instability can occur depends on the maximum allowed compactness of a GR-stable star, which in turn is highly EoS-dependent. In particular, for the polytropic EoS employed in this work, this would be $\beta_{0,\textrm{min}} \simeq 15$, as seen from Fig.~\ref{fig:inst1}. It is interesting to notice that $\beta_{0,\textrm{min}}$ can approach zero for some other physical, although somewhat artificial, systems, such as spherical shells of matter \cite{Lima2013b}.

In order to determine the actual observational signatures of this instability, it is essential to understand its nonlinear development. As discussed in the main text, only the linear piece of the coupling function $\alpha(\phi)$ is relevant for the linear stability analysis, but in order to solve the full nonlinear problem, we must specify its full form. In this work we focused on two representative coupling functions, displayed in Eqs.~(\ref{eq:alpha1}) and (\ref{eq:alpha2}). The first one (Model 1) approximates analytically the case of the more fundamental action (\ref{eq:S_NMC}), while the second one (Model 2) is a simple but popular model that truncates $\alpha(\phi)$ to linear order. 

The appearance of unstable scalar modes in the GR solution is accompanied by the appearance of additional equilibrium solutions with nonzero scalar charge. In theories with $\beta_0<0$, it is well known that, for any unstable GR solution, there is a stable, energetically favored scalarized solution with the same baryonic mass, to which the unstable configuration evolves (cf. Sec.~\ref{sec:betaminus}). Remarkably, the prediction that a given unstable star would undergo spontaneous scalarization relies only on the value of $\beta_0$, although the properties of the final state do depend on the full form of the coupling function.

In striking contrast with the $\beta_0<0$ case, we find that the very outcome of the instability in theories with $\beta_0>0$ is sensitive to higher-order terms in the coupling function. Our results for the representative coupling functions (\ref{eq:alpha1}) and (\ref{eq:alpha2}) are described in Sec.~\ref{sec:betaplus}.
In particular, for Model 1 we find that scalarized configurations exist that are stable and energetically favored over \textit{some} unstable GR solutions, and we numerically confirm that the former are the final state of the latter's evolution. For other unstable GR solutions (those with high enough mass), however, we show that there are no stable equilibrium solutions to which they can evolve, and that they eventually undergo gravitational collapse (cf.~Sec.~\ref{sec:M1}).
On the other hand, for Model 2, our numerical simulations corroborate previous results \cite{Palenzuela2016}, which found no evidence of spontaneous scalarization and that the end state of the instability would generically be collapse to a black hole. This is endorsed by an energy balance analysis of the existing equilibrium solutions, whereby we show that scalarized solutions in these theories have higher total mass than the GR ones (cf. Sec.~\ref{sec:M2}). 
Interestingly, we have verified that, if we consider coupling functions with higher-order polynomial terms in Eq.~(\ref{eq:a_taylor}), it is possible to interpolate between the qualitative behavior of Model 2 and Model 1 \cite{MathematicaNotebook}. 

Our results suggest observational signatures that could be searched for in order to probe STTs with $\beta_0>0$. In these theories, young neutron stars, which typically have lower masses, would be identical to the ones in GR (or very similar, if $\alpha_0 \neq 0$ but small). However, if these stars are formed in a material-rich environment, they will gradually increase their mass and compactness due to accretion and may eventually become sufficiently compact in order to develop unstable scalar modes. In Model 1 (or similar models), the unstable star is expected to spontaneously scalarize, giving rise to the characteristic observational imprints. These include changes in the orbital dynamics \cite{Damour1996}, in the redshift of surface atomic lines \cite{DeDeo2003}, and in the gravitational wave emission \cite{Novak1998} and spectrum \cite{Sotani2004}. Note that these results have been worked out for the $\beta_0<0$ case, but the details when $\beta_0>0$ still need to be investigated.
On the other hand, in Model 2 (or similar models), the unstable star is expected to undergo gravitational collapse. In this case, the mere observation of a stable neutron star above a certain critical compactness could be used to impose new constraints.
Additionally, highly massive neutron stars are possible outcomes of the coalescence and merger of binary neutron star systems, which are target sources for gravitational wave detectors such as Advanced LIGO. The effects discussed here for STTs with $\beta_0>0$ might be relevant both in the coalescence stage if an effect akin to dynamical scalarization \cite{Barausse2013,Shibata2014} takes place, and for the after-merger dynamics of the system.
We hope that these possibilities will be explored in future work.

\acknowledgments
The authors are glad to thank Eric Poisson, Luis Lehner, and Jonah Miller for helpful discussions. 
R.M.~acknowledges financial support from Conselho Nacional de Desenvolvimento Cient\'{i}fico e Tecnol\'{o}gico (CNPq). N.O.~acknowledges financial support from the CONACyT Grant No. 262714, and from the Perimeter Institute for Theoretical Physics. Research at the Perimeter Institute is supported by the Government of Canada through Industry Canada and by the Province of Ontario through the Ministry of Research and Innovation.

\appendix
\section{Numerical methods}

In this Appendix we describe the numerical techniques employed to solve the Cauchy problem for the scalar-tensor-Euler system consisting of Eqs.~(\ref{eq:G_eq}),~(\ref{eq:phi_eq}),~(\ref{eq:eom}), and~(\ref{eq:MB_conserv}), in a spacetime split according to the $3+1$ formalism, with initial data constrained by the system itself in the static limit. 
In spherical symmetry, the problem reduces to $1+1$ dimensions. The spatial radial coordinate $r$ is discretized using a uniform grid with $N_r + 1$ nodes $r_i = i\Delta r$, $i \in \{0, 1, 2, ..., N_r\}$, with $\Delta r := r_\textrm{max}/N_r$, where $r_\textrm{max} \in \Real^+$ is the outer boundary of the numerical domain. The time coordinate is also discretized in a uniform grid with $N_t + 1$ nodes $t^n = n\Delta t$, $n \in \{0, 1, 2, ..., N_t\}$, with $\Delta t := t_\textrm{max}/N_t$, where $t_\textrm{max} \in \Real^+$ is the maximum given coordinate time for a simulation.

\subsection{Initial data}

For all our simulations, the initial data consists of static equilibrium stars, described by solutions of Eqs.~(\ref{eq:dm})-(\ref{eq:dp}) subject to the boundary conditions (\ref{eq:bc}). We compute them numerically by a shooting-like method, as follows. First, we integrate Eqs.~(\ref{eq:dm})-(\ref{eq:dp}) from $r=0$ with conditions $m(0)=0$, $\phi(0)=\phi_c$, $\phi'(0)=0$, $\tilde{p}(0)=\tilde{p}_c$, and $N(0)=N_c$, where $\phi_c$ is a guessed value for the central field and $N_c$ is an arbitrary value that will be fixed \textit{a posteriori}. For the first grid points, we employ second-order Taylor expansions in order to regularize Eqs.~(\ref{eq:dm})-(\ref{eq:dp}) around $r=0$.
The integration is performed with a fourth-order accurate Runge-Kutta algorithm up to the surface of the star, $r=R_s$, defined by $\tilde{p}(R_s) = 0$, or equivalently, $\tilde{\rho}(R_s) = 0$. We refine the step size of the integrator until $\tilde{\rho}(R_s)/\tilde{\rho}_c < 10^{-11}$. With the quantities computed at $R_s$, we calculate the left-hand side of Eq.~(\ref{match}). We iterate the value of $\phi_c$ and repeat the above procedure until this quantity vanishes, within a numerical tolerance of the order $10^{-15}$. For $r>R_s$, we integrate Eqs.~(\ref{eq:dm})-(\ref{eq:dphi}) only, up to $r=r_\textrm{max}$. If $r_\textrm{max}$ is sufficiently large, we should have $N(r_\textrm{max})\approx 1/A(r_\textrm{max})$, corresponding to an asymptotically flat spacetime. To ensure that this condition is satisfied, we rescale $N(r)$ by  $1/[A(r_\textrm{max}) N(r_\textrm{max})]$.
We have verified that the initial data is second-order self-convergent.

\subsection{Evolution of the scalar-tensor-fluid system}

Given the nonlinear nature of the dynamic scalar-tensor-Euler system, it is expected to generically develop shocks and rarefaction waves, which imply the development of unbounded gradients in the fluid variables even starting from smooth initial data. Since the usual finite differences schemes assume certain smoothness of the solution, they fail to handle fluid shocks. On the other hand, finite volume schemes together with high resolution shock capturing (HRSC) methods are alternative approaches which can consistently deal with discontinuous solutions. They are based on the discretization of the evolution equations in their integral flux-conservative form, which thus requires the definition of conservative variables to be evolved in a mesh of $N_r$ finite volume cells, the $i$th cell centered at $r_i - \Delta r/2$, $i \in \{ 1, 2, ..., N_r\}$ (see, for instance, Ref.~\cite{LeVeque-Book_FV}). The particularities of a HRSC method depend on the way it reconstructs primitive variables on the cell interfaces and then solves the Riemann problem arising at each interface. We have implemented a HRSC method based on the standard Harten-Lax-van Leer-Einfeldt approximate Riemann solver with a linear piecewise reconstructor of variables. For technical details on this method, we refer the reader to the specialized literature, Ref.~\cite{fBjFjIjMjM97} for instance (see also Ref.~\cite{fGfLmM12} for a revisit of the spherically symmetric case and a description of the usual regularization of the coordinate singularity at $r=0$). 

Using this method, we solve the fluid-scalar-field evolution system written in the hyperbolic flux-conservative form~(\ref{eq:flux-conservative-form}) with ${\bf q} = ( \tilde{D}, \tilde{S}, \tilde{\tau} , \eta, \psi )^T$, and ${\bf F}$ and ${\bf S}$ given by Eqs.~(\ref{eq:F_D})-(\ref{eq:S_psi}). For the time evolution, we employ a third-order accurate Runge-Kutta integrator, which guarantees that the Total Variation Diminishing condition is satisfied \cite{LeVeque-Book_FV}. No artificial dissipation was needed. For all the results presented in this work, we use a typical spatial resolution $\Delta r$ of the order of $10^{-4}r_\textrm{max}$, while the time step size $\Delta t$ is constrained by the Courant-Friedrichs-Lewy condition $\Delta t/\Delta x = \kappa_\textrm{CFL} < 1$, which we ensure by setting $\kappa_\textrm{CFL} = 0.25$.

Concerning the evolution of the metric functions, we discretize Eq.~(\ref{eq:momentum_constraint}) on the original grid $(t^n,r_i)$, and integrate it simultaneously with the fluid, using the same third-order Runge-Kutta algorithm. This determines the metric function $A(t^n,r_i)$ through the relation $A(t^n,r_i) = \left[ 1 - 2m(t^n,r_i)/r_i \right]^{-1/2}$. In turn, the lapse function $N(t^n,r_i)$ is updated at every time step by integrating Eq.~(\ref{eq:lapse_condition}) in space by means of a second-order quadrature scheme. Whenever needed, values of the metric functions $N$ and $A$ at the finite volume cell centers are interpolated by a second-order Lagrange polynomial. 

\subsubsection{Boundary conditions}

At the inner boundary, given by $r = 0$, we impose the usual regularity conditions due to the spherical symmetry of the problem: the quantities $\tilde{D}$, $\tilde{\tau}$ and $\psi$ are even functions of $r$ at all times, whereas $\tilde{S}$ and $\eta$ are odd functions. We implement these conditions through second-order Lagrange interpolations using two ghost cells at the left edge of the domain.

At the outer boundary of the domain, at $r=r_\textrm{max}$, we demand an outgoing flow condition for the fluid, meaning $\left. \partial_r {\bf q} \right|_{r_\textrm{max}} = 0$ at all times, whereas for the scalar field we impose~ \cite{Novak1998b}
\begin{equation}\label{eq:outgoing_condition}
\left.\left( A\eta + A\psi + \frac{\phi-\phi_0}{r} \right)\right|_{r_\textrm{max}} = 0,
\end{equation}
which follows from the outgoing wave condition $\phi(t,r)  \overset{r \to \infty}{\longrightarrow} \phi_0 + F(t-r)/r$, where $F$ is an arbitrary function of the retarded time. Note that Eq. (\ref{eq:outgoing_condition}) is strictly valid only in a flat spacetime, but becomes a good approximation when $r_\textrm{max}$ is sufficiently large, and the field amplitude is small. In particular, we have chosen $r_\textrm{max}$ such that $r_\textrm{max}/R_s > 20$, in which case the field amplitude $\phi(r_\textrm{max})$ is typically of order $10^{-5}$.

\subsubsection{Vacuum region}

Outside of the star, the fluid variables should in principle vanish, but a special treatment is required when HRSC methods are employed. These methods are generically unable to handle the vacuum scenario because, in the variable reconstruction stage, the relation between primitive and conservative variables must be inverted, and this transformation becomes singular in vacuum. In order to alleviate this issue, we have implemented the standard {\it atmosphere} artifice, which consists in setting up {\it ad hoc} a constant, very small baryon density outside the star at all times, so that the HRSC method still works in the exterior, while the effects on the stellar dynamics are negligible. We follow the usual implementation described, e.g., in Sec.~VI of Ref.~\cite{jFmMwSmT00}, with an atmosphere density $\rho_\textrm{atm}$ of the order of $10^{-11}\tilde{\rho}_c$, where $\tilde{\rho}_c$ is a given stellar central density, and a criterion to reset the density to the atmosphere value whenever it drops below $f_\textrm{thres} \rho_\textrm{atm}$, where $f_\textrm{thres}\sim 10^2-10^4$.

\begin{figure}[tb]
\includegraphics[width=7.5cm]{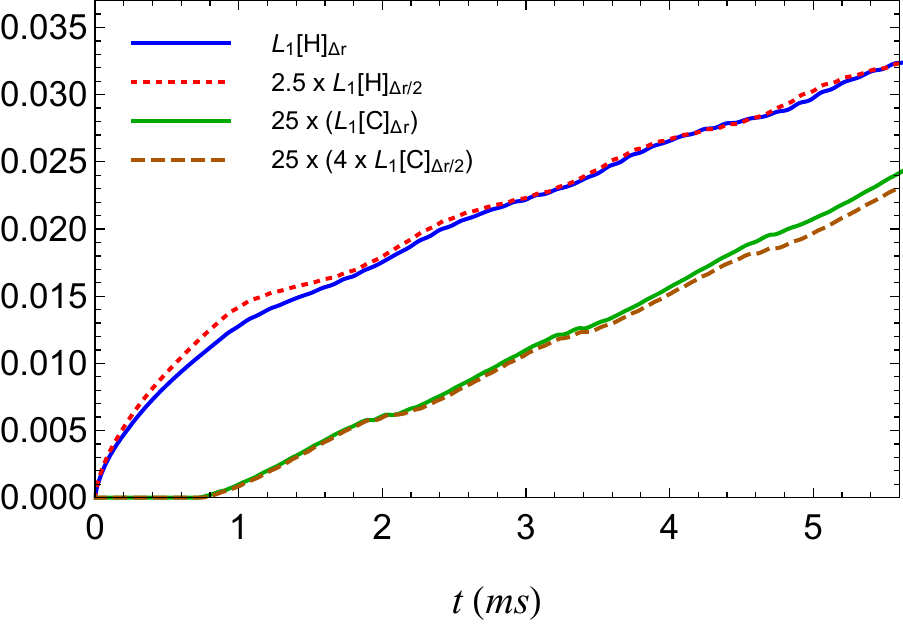}
\caption{$L_1$-norm of $\mathcal{H}$ and $\mathcal{C}$ as a function of time, for the evolution of solution 1 in Table \ref{table:initialdata2}, for two consecutive resolutions, $\Delta r = 10^{-4} r_\textrm{max}$ and $\Delta r/2$. We multiply the higher resolution curves by a factor that makes them roughly superpose on the lower resolution ones. For the Hamiltonian constraint, this factor is around 2.5, indicating self-convergence to order $\gtrsim 1.5$. For the wave equation constraint, the factor is around 4, indicating second-order convergence. For the sake of visualization, we rescale the $L_1[\mathcal{C}]$ curves by a factor of 25.}
\label{fig:convergence}
\end{figure}

\subsection{Convergence tests}

In order to check the self-convergence of the numerical solutions, it is useful to monitor the preservation of the Hamiltonian constraint $\mathcal{H} = 0$ and the scalar field wave equation constraint $\mathcal{C} = 0$ at each time step, where
\begin{eqnarray}
{\cal H} &:=& \partial_r m - \frac{r^2}{2} \left[ \eta^2 + \psi^2 + 8\pi a^4 (\tilde{\tau} + \tilde{D} )\right], \\
{\cal C} &:=&  \partial _r \phi - A \eta.
\end{eqnarray}
A usual test consists in evaluating the $L_1$-norm of the deviations of the constraints from zero at each time step, where $L_1[f] := \int_0^{r_\textrm{max}} |f(r)| dr$, and then comparing this quantity for two consecutive resolutions $\Delta r$ and $\Delta r/2$. A self-convergence factor $\kappa$ can be defined as $\kappa := L_1[f]_{\Delta r}/L_1[f]_{\Delta r/2}$, where $f$ is either ${\cal H}$ or ${\cal C}$. The solution is said to be self-convergent if $\kappa > 1$ and  the order of convergence is $\sqrt{\kappa}$. 
For every simulation presented here, except when gravitational collapse is imminent, we have made sure that our numerical solutions are indeed self-convergent to a constrained solution of the scalar-tensor-Euler system. In particular, in Fig.~\ref{fig:convergence}, we show $L_1[{\cal H}]$ and $L_1[{\cal C}]$  for two consecutive resolutions in the case of spontaneous scalarization within Model 1 with $\beta=100$ (see Fig.~\ref{fig:betaplusEvol1}). From the figure we infer $\kappa \simeq 2.5$ for the Hamiltonian constraint and $\kappa \simeq 4$ for the wave equation constraint.

\begin{figure}[b]
\includegraphics[width=8.5cm]{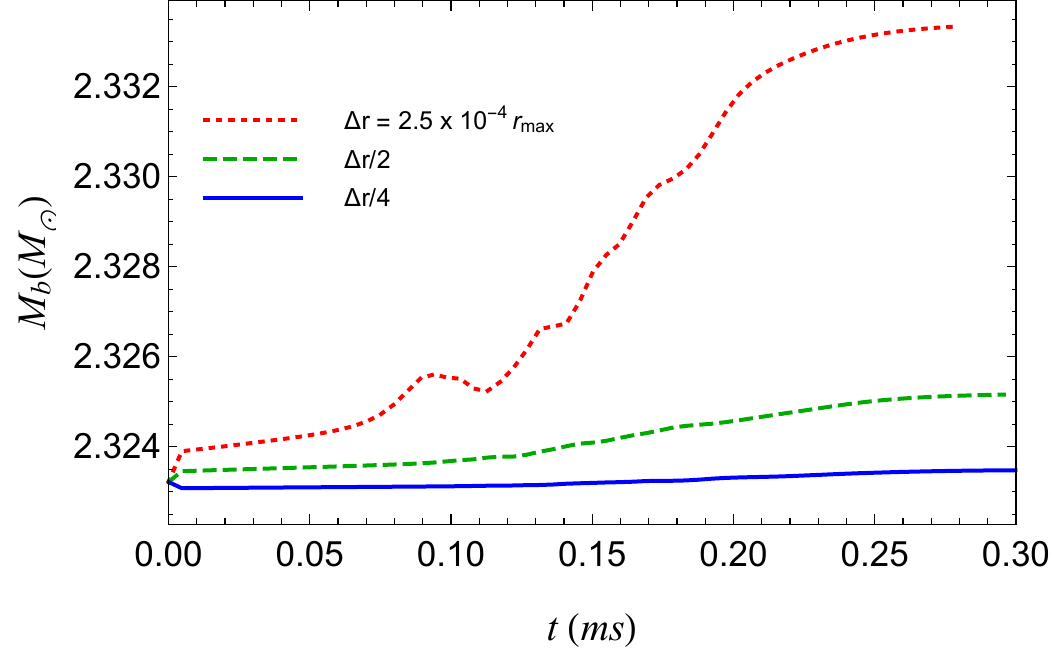}
\caption{Total baryonic mass as a function of time for three consecutive numerical resolutions in the case of gravitational collapse of initial data 6 in Table~\ref{table:initialdata3}. We see the baryonic mass trending to a constant value as the resolution is increased, as expected.}
\label{fig:baryon_mass}
\end{figure}

For initial data consisting of hydrodynamically stable GR solutions with unstable scalar modes, a complementary convergence test consists in the direct comparison of the rate of exponential growth of the scalar field obtained numerically, against the one predicted by the linear analysis. We infer the growth rate from the numerical data by computing $\Omega = (\ln \phi_c^{(b)} - \ln \phi_c^{(a)})/(t^{(b)} - t^{(a)})$, where $t^{(a)}$ is the time such that $\phi_c^{(a)} := |\phi_c(t^{(a)}) - \phi_0| = 10^{-14}$ and $t^{(b)}$ is such that $\phi_c^{(b)} := |\phi_c(t^{(b)}) - \phi_0| = 10^{-3}$. 
Let $\Delta \Omega_q$ be the absolute difference of the predicted growth rate and the one obtained numerically with a spatial resolution $\Delta r/q$, $q \in \{1,2\}$, where $\Delta r = 10^{-4} r_\textrm{max}$. For the same case discussed above (see Figs.~\ref{fig:betaplusEvol1} and \ref{fig:convergence}), we have $\Delta \Omega_1 = 0.0818$ and $\Delta \Omega_2 = 0.0167$. Therefore, the convergence factor is $\Delta \Omega_1 / \Delta \Omega_2 \sim 4.9$, indicating convergence to order $\gtrsim 2$.

Yet another check of our numerical machinery consists in monitoring the baryonic mass, defined in Eq.~(\ref{eq:Total_BM}), which should be conserved in time due to Eq.~(\ref{eq:MB_conserv}). For the particular case of gravitational collapse of initial data 6 in Table \ref{table:initialdata3} (see also Fig.~\ref{fig:betaplusEvol3}), 
in Fig.~\ref{fig:baryon_mass} we show the trend of the total baryonic mass approaching a constant value as we increase the grid resolution. In particular, for the highest resolution, we have a variation of the baryonic mass of less than 1 part in $10^5$.

\bibliographystyle{unsrt}
\bibliography{references}

\begin{thebibliography}{10}

\bibitem{Damour1992}
T.~Damour and G.~Esposito-Far\`ese.
\newblock Tensor-multi-scalar theories of gravitation.
\newblock {\em Classical Quantum Gravity}, 9:2093, 1992.

\bibitem{Fujii2003}
Y.~Fujii and K.~Maeda.
\newblock {\em The Scalar-Tensor Theory of Gravitation}.
\newblock Cambridge University Press, Cambridge, England, 2003.

\bibitem{Salgado2006}
M.~Salgado.
\newblock {The Cauchy problem of scalar-tensor theories of gravity}.
\newblock {\em Class. Quantum Gravity}, 23:4719, 2006.

\bibitem{Faraoni2004}
V.~Faraoni.
\newblock {\em Cosmology in Scalar-Tensor Gravity}.
\newblock Springer, New York, 2004.

\bibitem{Clifton2012}
T.~Clifton, P.~G. Ferreira, A.~Padilla, and C.~Skordis.
\newblock {Modified gravity and cosmology}.
\newblock {\em Physics Reports}, 513:1, 2012.

\bibitem{Horbatsch2015}
M.~Horbatsch, H.~O. Silva, D.~Gerosa, P.~Pani, E.~Berti, L.~Gualtieri, and
  U.~Sperhake.
\newblock {Tensor-multi-scalar theories: relativistic stars and $3 + 1$
  decomposition}.
\newblock {\em Class. Quantum Gravity}, 32:204001, 2015.

\bibitem{Ramazanoglu2016}
F.~M. Ramazanoglu and F.~Pretorius.
\newblock {Spontaneous scalarization with massive fields}.
\newblock {\em Phys. Rev. D}, 93:064005, 2016.

\bibitem{Damour1993a}
T.~Damour and K.~Nordtvedt.
\newblock {Tensor-scalar cosmological models and their relaxation toward
  general relativity}.
\newblock {\em Phys. Rev. D}, 48:3436, 1993.

\bibitem{Damour1993b}
T.~Damour and K.~Nordtvedt.
\newblock {General relativity as a cosmological attractor of tensor-scalar
  theories}.
\newblock {\em Phys. Rev. Lett.}, 70:2217, 1993.

\bibitem{Jordan1959}
R.~Jordan.
\newblock Zum gegenwartigen stand der diracschen kosmologischen hypothesen.
\newblock {\em Zeitschrift fur Phys.}, 157:112, 1959.

\bibitem{Brans1961}
C.~Brans and R.~H. Dicke.
\newblock {Mach's Principle and a Relativistic Theory of Gravitation}.
\newblock {\em Phys. Rev.}, 124:925, 1961.

\bibitem{Bertotti2003}
B.~Bertotti, L.~Iess, and P.~Tortora.
\newblock {A test of general relativity using radio links with the Cassini
  spacecraft}.
\newblock {\em Nature}, 425:374, 2003.

\bibitem{Damour1996a}
T.~Damour and G.~Esposito-Far{\`{e}}se.
\newblock {Testing gravity to second post-Newtonian order: A field-theory
  approach}.
\newblock {\em Phys. Rev. D}, 53:5541, 1996.

\bibitem{Damour1993}
T.~Damour and G.~Esposito-Far\`ese.
\newblock Nonperturbative strong-field effects in tensor-scalar theories of
  gravitation.
\newblock {\em Phys. Rev. Lett.}, 70:2220, 1993.

\bibitem{Damour1996}
T.~Damour and G.~Esposito-Far\`ese.
\newblock Tensor-scalar gravity and binary-pulsar experiments.
\newblock {\em Phys. Rev. D}, 54:1474, 1996.

\bibitem{Salgado1998}
M.~Salgado, D.~Sudarsky, and U.~Nucamendi.
\newblock Spontaneous scalarization.
\newblock {\em Phys. Rev. D}, 58:124003, 1998.

\bibitem{Harada1998}
T.~Harada.
\newblock Neutron stars in scalar-tensor theories of gravity and catastrophe
  theory.
\newblock {\em Phys. Rev. D}, 57:4802, 1998.

\bibitem{Freire2012}
P.~C.~C. Freire, N.~Wex, G.~Esposito-Far\`ese, J.~P.~W. Verbiest, M.~Bailes,
  B.~A. Jacoby, M.~Kramer, I.~H. Stairs, J.~Antoniadis, and G.~H. Janssen.
\newblock {The relativistic pulsar-white dwarf binary PSR J1738+0333 - II. The
  most stringent test of scalar-tensor gravity}.
\newblock {\em Mon. Not. R. Astron. Soc.}, 423:3328, 2012.

\bibitem{Abbott2016a}
B.~P.~Abbott et~al. (LIGO Scientific~Collaboration and Virgo Collaboration).
\newblock {Observation of Gravitational Waves from a Binary Black Hole Merger}.
\newblock {\em Phys. Rev. Lett.}, 116:061102, 2016.

\bibitem{Barausse2013}
E.~Barausse, C.~Palenzuela, M.~Ponce, and L.~Lehner.
\newblock Neutron-star mergers in scalar-tensor theories of gravity.
\newblock {\em Phys. Rev. D}, 87:081506, 2013.

\bibitem{Shibata2014}
M.~Shibata, K.~Taniguchi, H.~Okawa, and A.~Buonanno.
\newblock {Coalescence of binary neutron stars in a scalar-tensor theory of
  gravity}.
\newblock {\em Phys. Rev. D}, 89:084005, 2014.

\bibitem{Palenzuela2014}
C.~Palenzuela, E.~Barausse, M.~Ponce, and L.~Lehner.
\newblock {Dynamical scalarization of neutron stars in scalar-tensor gravity
  theories}.
\newblock {\em Phys. Rev. D}, 89:044024, 2014.

\bibitem{Sampson2014}
L.~Sampson, N.~Yunes, N.~Cornish, M.~Ponce, E.~Barausse, A.~Klein,
  C.~Palenzuela, and L.~Lehner.
\newblock {Projected constraints on scalarization with gravitational waves from
  neutron star binaries}.
\newblock {\em Phys. Rev. D}, 90:124091, 2014.

\bibitem{Berti2015a}
E.~Berti, E.~Barausse, V.~Cardoso, L.~Gualtieri, P.~Pani, U.~Sperhake, L.~C.
  Stein, N.~Wex, K.~Yagi, T.~Baker, C.~P. Burgess, F.~S. Coelho, D.~Doneva,
  A.~De Felice, P.~G. Ferreira, P.~C.~C. Freire, J.~Healy, C.~Herdeiro,
  M.~Horbatsch, B.~Kleihaus, A.~Klein, K.~Kokkotas, J.~Kunz, P.~Laguna, R.~N.
  Lang, T.~G.~F. Li, T.~Littenberg, A.~Matas, S.~Mirshekari, H.~Okawa, E.~Radu,
  R.~O'Shaughnessy, B.~S. Sathyaprakash, C.~V.~D. Broeck, H.~A. Winther,
  H.~Witek, M.~E. Aghili, J.~Alsing, B.~Bolen, L.~Bombelli, S.~Caudill,
  L.~Chen, J.~C. Degollado, R.~Fujita, C.~Gao, D.~Gerosa, S.~Kamali, H.~O.
  Silva, J.~G. Rosa, L.~Sadeghian, M.~Sampaio, H.~Sotani, and M.~Zilhao.
\newblock {Testing general relativity with present and future astrophysical
  observations}.
\newblock {\em Class. Quantum Gravity}, 32:243001, 2015.

\bibitem{Mendes2015}
R.~F.~P. Mendes.
\newblock Possibility of setting a new constraint to scalar-tensor theories.
\newblock {\em Phys. Rev. D}, 91:064024, 2015.

\bibitem{Palenzuela2016}
C.~Palenzuela and S.~Liebling.
\newblock Constraining scalar-tensor theories of gravity from the most massive
  neutron stars.
\newblock {\em Phys. Rev. D}, 93:044009, 2016.

\bibitem{Flanagan2004}
{\'{E}}.~{\'{E}}. Flanagan.
\newblock {The conformal frame freedom in theories of gravitation}.
\newblock {\em Class. Quantum Gravity}, 21:3817, 2004.

\bibitem{Sotiriou2008}
T.~P. Sotiriou, S.~Liberati, and V.~Faraoni.
\newblock {Theory of gravitation theories: a no-progress report}.
\newblock {\em Int. J. Mod. Phys. D}, 17:399, 2008.

\bibitem{Alcubierre-Book}
M.~Alcubierre.
\newblock {\em Introduction to 3 + 1 Numerical Relativity}.
\newblock Oxford University Press, New York, 2008.

\bibitem{LeVeque-Book_FD}
R.~J. LeVeque.
\newblock {\em Finite difference methods for ordinary and partial differential
  equations : steady-state and time-dependent problems}.
\newblock Society for Industrial and Applied Mathematics, Philadelphia, PA,
  USA, 2007.

\bibitem{LeVeque-Book_FV}
R.~J. LeVeque.
\newblock {\em Finite Volume Methods for Hyperbolic Problems}.
\newblock Cambridge University Press, Cambridge, England, 2002.

\bibitem{Novak2000}
J.~Novak and J.~M. Ib{\'{a}}{\~{n}}ez.
\newblock {Gravitational Waves from the Collapse and Bounce of a Stellar Core
  in Tensor-Scalar Gravity}.
\newblock {\em Astrophys. J.}, 533:392, 2000.

\bibitem{Gerosa2016}
D.~Gerosa, U.~Sperhake, and C.~D. Ott.
\newblock {Numerical simulations of stellar collapse in scalar-tensor theories
  of gravity}.
\newblock arXiv:1602.06952.

\bibitem{Coquereaux}
R.~Coquereaux and G.~Esposito-Far\`ese.
\newblock {The theory of Kaluza-Klein-Jordan-Thiry revisited}.
\newblock {\em Annales de l'I. H. P.}, 52:113, 1990.

\bibitem{Haensel2007}
P.~Haensel, A.~Y. Potekhin, and D.~G. Yakovlev.
\newblock {\em Neutron stars 1: Equation of state and structure}.
\newblock Springer, New York, 2007.

\bibitem{Read2009}
J.~Read, B.~Lackey, B.~Owen, and J.~L. Friedman.
\newblock {Constraints on a phenomenologically parametrized neutron-star
  equation of state}.
\newblock {\em Phys. Rev. D}, 79:124032, 2009.

\bibitem{Antoniadis2013}
J.~Antoniadis, P.~C.~C. Freire, N.~Wex, T.~M. Tauris, R.~S. Lynch, M.~H. van
  Kerkwijk, M.~Kramer, C.~Bassa, V.~S. Dhillon, T.~Driebe, J.~W.~T. Hessels,
  V.~M. Kaspi, V.~I. Kondratiev, N.~Langer, T.~R. Marsh, M.~A. McLaughlin,
  T.~T. Pennucci, S.~M. Ransom, I.~H. Stairs, J.~van Leeuwen, J.~P.~W.
  Verbiest, and D.~G. Whelan.
\newblock {A massive pulsar in a compact relativistic binary}.
\newblock {\em Science}, 340:1233232, 2013.

\bibitem{Ozel2016}
F.~\"{O}zel and P.~Freire.
\newblock {Masses, Radii, and Equation of State of Neutron Stars}.
\newblock {\em Annu. Rev. Astron. Astrophys.}, 54, 2016.

\bibitem{Douchin2001}
F.~Douchin and P.~Haensel.
\newblock A unified equation of state of dense matter and neutron star
  structure.
\newblock {\em Astron. Astrophys}, 380:151, 2001.

\bibitem{Birrell1982}
N.~D. Birrell and P.~C.~W. Davies.
\newblock {\em Quantum Fields in Curved Space}.
\newblock Cambridge University Press, Cambridge, England, 1982.

\bibitem{Harada1997}
T.~Harada.
\newblock Stability analysis of spherically symmetric star in scalar-tensor
  theories of gravity.
\newblock {\em Prog. Theor. Phys.}, 98:359, 1997.

\bibitem{Sotani2005}
H.~Sotani and K.~D. Kokkotas.
\newblock {Stellar oscillations in scalar-tensor theory of gravity}.
\newblock {\em Phys. Rev. D}, 71:124038, 2005.

\bibitem{Sotani2014}
H.~Sotani.
\newblock {Scalar gravitational waves from relativistic stars in scalar-tensor
  gravity}.
\newblock {\em Phys. Rev. D}, 89:064031, 2014.

\bibitem{Lima2010}
W.~C.~C. Lima and D.~A.~T. Vanzella.
\newblock Gravity-induced vacuum dominance.
\newblock {\em Phys. Rev. Lett.}, 104:161102, 2010.

\bibitem{Lima2010b}
W.~C.~C. Lima, G.~E.~A. Matsas, and D.~A.~T. Vanzella.
\newblock Awaking the vacuum in relativistic stars.
\newblock {\em Phys. Rev. Lett.}, 105:151102, 2010.

\bibitem{Mendes2014}
R.~F.~P. Mendes, G.~E.~A. Matsas, and D.~A.~T. Vanzella.
\newblock {Quantum versus classical instability of scalar fields in curved
  backgrounds}.
\newblock {\em Phys. Rev. D}, 89:047503, 2014.

\bibitem{Novak1998}
J.~Novak.
\newblock Neutron star transition to a strong-scalar-field state in
  tensor-scalar gravity.
\newblock {\em Phys. Rev. D}, 58:064019, 1998.

\bibitem{Alcubierre2010}
M.~Alcubierre, J.~C. Degollado, D.~N{\'{u}}{\~{n}}ez, M.~Ruiz, and M.~Salgado.
\newblock Dynamic transition to spontaneous scalarization in boson stars.
\newblock {\em Phys. Rev. D}, 81:124018, 2010.

\bibitem{Ruiz2012}
M.~Ruiz, J.~C. Degollado, M.~Alcubierre, D.~N{\'{u}}{\~{n}}ez, and M.~Salgado.
\newblock {Induced scalarization in boson stars and scalar gravitational
  radiation}.
\newblock {\em Phys. Rev. D}, 86:104044, 2012.

\bibitem{Pani2011}
P.~Pani, V.~Cardoso, E.~Berti, J.~Read, and M.~Salgado.
\newblock Vacuum revealed: The final state of vacuum instabilities in compact
  stars.
\newblock {\em Phys. Rev. D}, 83:081501, 2011.

\bibitem{Baumgarte2010}
T.~W. Baumgarte and S.~L. Shapiro.
\newblock {\em Numerical Relativity}.
\newblock Cambridge University Press, 2010.

\bibitem{MathematicaNotebook}
R.~F.~P. Mendes and N.~Ortiz.
\newblock {A} {M}athematica [{W}olfram {R}esearch, {I}nc., {M}athematica 10.2,
  {C}hampaign, {IL} (2015)] notebook containing routines to construct
  equilibrium solutions in {STT}s and complementary plots to this paper is
  available at the public repository
  \url{https://bitbucket.org/nestor_ortiz/highly-compact-neutron-stars-in-stts.git}.

\bibitem{Lima2013b}
W.~C.~C. Lima, R.~F.~P. Mendes, G.~E.~A. Matsas, and D.~A.~T. Vanzella.
\newblock {Awaking the vacuum with spheroidal shells}.
\newblock {\em Phys. Rev. D}, 87:104039, 2013.

\bibitem{DeDeo2003}
S.~DeDeo and D.~Psaltis.
\newblock {Towards New Tests of Strong-Field Gravity with Measurements of
  Surface Atomic Line Redshifts from Neutron Stars}.
\newblock {\em Phys. Rev. Lett.}, 90:141101, 2003.

\bibitem{Sotani2004}
H.~Sotani and K.~D. Kokkotas.
\newblock {Probing strong-field scalar-tensor gravity with gravitational wave
  asteroseismology}.
\newblock {\em Phys. Rev. D}, 70:084026, 2004.

\bibitem{fBjFjIjMjM97}
F.~Banyuls, J.~A. Font, J.~M. Ib{\'{a}}{\~{n}}ez, J.~M. Mart\'in, and J.~A.
  Miralles.
\newblock Numerical $3+1$ general relativistic hydrodynamics: A local
  characteristic approach.
\newblock {\em The Astrophysical Journal}, 476:221, 1997.

\bibitem{fGfLmM12}
F.S. Guzm\'an, F.D. Lora-Clavijo, and M.D. Morales.
\newblock {Revisiting spherically symmetric relativistic hydrodynamics}.
\newblock {\em {Revista mexicana de f\'isica E}}, 58:84, 12 2012.

\bibitem{Novak1998b}
J.~Novak.
\newblock {Spherical neutron star collapse toward a black hole in a
  tensor-scalar theory of gravity}.
\newblock {\em Phys. Rev. D}, 57:4789, 1998.

\bibitem{jFmMwSmT00}
J.~A. Font, M.~Miller, W.~Suen, and M.~Tobias.
\newblock Three-dimensional numerical general relativistic hydrodynamics:
  Formulations, methods, and code tests.
\newblock {\em Phys. Rev. D}, 61:044011, 2000.

\end{thebibliography}

\end{document}